\documentclass[aps,pre,
superscriptaddress,superscriptaddress,showpacs,twocolumn]{revtex4-2}
\usepackage{graphicx,epsf}
\usepackage{hyperref}
\usepackage{xcolor}
\usepackage{psfrag}
\usepackage{amsmath}
\usepackage{amssymb}
\usepackage{orcidlink}

\begin{document}

\title{Shape of semiflexible polymer chains:  L\'evy walks vs. Persistent random walks}

\author{V. Blavatska\orcidlink{0000-0001-6158-1636}}\email{blavatskav@gmail.com}
\affiliation{Yukhnovskii Institute for Condensed
Matter Physics of the National Academy of Sciences of Ukraine,\\
1 Svientsitskii Str., UA-79011 Lviv, Ukraine}

\author{Yu. Honchar\orcidlink{0000-0003-2660-4593}}
\affiliation{Yukhnovskii Institute for Condensed
Matter Physics of the National Academy of Sciences of Ukraine,\\
1 Svientsitskii Str., UA-79011 Lviv, Ukraine}
\affiliation{ ${\mathbb L}^4$ Collaboration $\&$ Doctoral College for the Statistical Physics of Complex Systems, Leipzig-Lorraine-Lviv-Coventry, Europe}
\author{Yu. Holovatch\orcidlink{0000-0002-1125-2532}}
\affiliation{Yukhnovskii Institute for Condensed
Matter Physics of the National Academy of Sciences of Ukraine,\\
1 Svientsitskii Str., UA-79011 Lviv, Ukraine}
\affiliation{ ${\mathbb L}^4$ Collaboration $\&$ Doctoral College for the Statistical Physics of Complex Systems, Leipzig-Lorraine-Lviv-Coventry, Europe}
\affiliation{Centre for Fluid and Complex Systems, Coventry University, Coventry CV1 5FB, UK}
\affiliation{Complexity Science Hub, 1030 Vienna, Austria}

\begin{abstract}
The conformational size and shape characteristics of semiflexible polymer chains, modeled via two distinct frameworks: persistent random walks (PRW) and  Lévy walks (LW), are analyzed through numerical simulations. We introduce a mapping that treats the mean jump length $\langle \ell \rangle$ of a L\'evy distribution ($\sim \ell^{-(1+\sigma)}$) as an effective persistence length  of the semiflexible polymer chain, controlled by the exponent $\sigma $.
While the conformational characteristics  of the PRW model statistically exhibit a sharp crossover from flexible coil to the rigid-rod-like behavior in the immediate vicinity of the maximum persistence probability $p \to 1$, the LW  model shows a much richer conformational behavior. 
 We obtain the quantitative estimates for universal shape parameters, such as asphericity \(\langle A_3 \rangle\) and prolateness \(\langle S \rangle\), which gradually increase with decreasing parameter $\sigma  $ (and thus with an increase of the persistence length), thus supporting previously observed findings for semiflexible polymers.  Our results demonstrate the effictiveness of applying L\'evy-walk statistics to capture the universal conformational properties of semiflexible polymer structures.
\end{abstract}
\date{\today}
\maketitle

\section{Introduction}

Random walks (RW) constitute one of the central paradigms of statistical physics by describing motion as a sequence of random steps \cite{Pearson1905}. They provide, for example, a way to model biological movement and transport on networks \cite{Codling2008,Masuda2017}. For unbiased, independent steps with finite variance, the mean-square displacement grows linearly with the number of steps. When each step takes equal time, this gives for the mean-square displacement the Brownian diffusion law
\cite{Einstein1905}:
\begin{equation}
\langle R^{2}(t)\rangle \sim t . \label{flexr}
\end{equation}

This behavior follows from the central limit theorem and forms the basis of the conventional theory of diffusion. Many complex systems, however, exhibit transport properties that deviate substantially from Brownian behavior. Such systems display anomalous diffusion, characterized by a nonlinear growth of the mean-square displacement \cite{Levy1937, BouchaudGeorges1990, Metzler2000, Viswanathan1999, Viswanathan2000, Golding2006, Campeau2022}: 
\begin{equation}
	\langle R^{2}(t)\rangle \sim t^{\nu}, \, \,  {\rm with} \, \, \nu\neq1. 
    \label{nuexp}
\end{equation}
Of particular interest is the superdiffusive regime, $\nu>1$, which emerges when rare but exceptionally long displacements dominate the dynamics. A natural description of such processes is provided by L\'evy statistics, in which the distribution of step lengths $\ell$ possesses a heavy power-law tail,
\begin{equation}
	P(\ell)\sim \ell^{-(1+\sigma)},
	\qquad 0<\sigma\leq2,
	\label{PLevy}
\end{equation}
leading to $\nu=1/\sigma$ in Eq. (\ref{nuexp}) (see e.g. \cite{Chechkin2006}).
While idealized L\'evy flights allow instantaneous, arbitrarily long jumps, physically realizable systems are more naturally described by L\'evy walks, in which the propagation velocity remains finite and long displacements require correspondingly long travel times. In a L\'evy walk, the discrete time required to make a jump of a given size is taken to be proportional to the jump size \cite{Dybiec2017,Palyulin2019,Bnichou2011}. This modification preserves the essential heavy-tailed character of the motion while maintaining physical consistency. As a result, L\'evy walks have become a standard model of superdiffusive transport in complex systems.

In contrast to Gaussian random walks, L\'evy processes generate trajectories containing occasional long jumps and lead to fundamentally different scaling properties. They have been observed or invoked in diverse contexts, including turbulent transport, animal foraging, bacterial motion, transport in heterogeneous media, and various non-equilibrium systems \cite{Solomon1993,Viswanathan1999,Ariel2015,Barthelemy2008,Brockmann2006,MantegnaStanley1995}.

In this paper we intend to discuss the above reasonings in the context of polymer physics. {
The  RW model has been successfully applied to describe flexible polymer molecules in a so-called $\Theta$ solvent (Gaussian polymers) \cite{desCloiseaux1991,Freed1985}. Within this framework, each consecutive step of the RW trajectory represents a single monomer of the polymer chain, so that the propagation time $t$ in Eq.~(\ref{flexr}) is equal to the number of monomers $N$ in the corresponding macromolecule. In the present study, we focus only on Gaussian polymer structures, neglecting excluded-volume effects arising from mutual-avoidance interactions between monomers. Despite its simplicity,  RW model perfectly  captures the universal conformational properties of \textit{flexible} polymers. In particular, for a size measure such as the  mean-square displacement of an $N$-step RW, the relation in Eq.~(\ref{flexr}) holds.

Many synthetic polymers, such as polyethylene, possess simple carbon backbones with monomers that can rotate around single bonds, making the chains highly flexible. On the other hand, complex organic macromolecules such as DNA, collagen, proteins \cite{Bustamante1994,Witz2008,Marko1995,Biswas2021,Gerrits2021} contain bulky repeating units with complicated molecular interactions that restrict bond rotation and are typical representatives of a class of \textit{semiflexible} polymers. Theoretically, such polymers are described within the framework of the worm-like-chain (or Kratky--Porod polymer-chain) model \cite{Kratky1949,doi1988theory}.
The key structural characteristic here is the persistence length $\ell_{\rm p}$, which is proportional to the bending rigidity of the chain. In the range $\ell_{\rm p}\ll N$, macromolecules statistically behave as flexible chains with an effective monomer size of the order of $\ell_{\rm p}$ and obey the scaling relation in Gaussian regime
\begin{equation}
	\langle R^2 \rangle \sim {N}{\ell_{\rm p}}.
	\label{semiflexr}
\end{equation}
On the other hand, for the case of extremely large values of $l_{\rm p}$ ($N\approx \ell_{\rm p}$), the polymer approaches the limit of a completely extended rigid rod with
\begin{equation}
	\langle R^2\rangle \sim N^{2}.
\end{equation}

To take the bending rigidity of semiflexible polymers into account, the model of biased, or persistent, random walks (PRW) is used \cite{Halley1985,Masoliver1993,Haydukivska2015}. In this model, higher probability $p$ is introduced for ``trans'' steps (taken in the same direction as the previous step) as compared to  ``gauche'' steps (leading to bending of the trajectory).
The constructed trajectories thus constitute the sequences of rigid segments of average length $\ell_{\rm p} \sim p$ with random relative orientations, where each segment, in turn, comprises a number of monomers proportional to $\ell_{\rm p}$. 

Our idea is to capture the physics of semiflexible polymers using a L\'evy walk (LW) model. In this representation, the trajectory is viewed as a sequence of straight segments with lengths $\ell$ drawn from the L\'evy distribution in Eq.~(\ref{PLevy}). The propagation time along each straight segment is equal to the number of monomers $\ell$ it comprises. The average jump length $\ell_{{\rm LW}}$, evaluated according to the distribution in Eq.~(\ref{PLevy}), is therefore directly related to the concept of the persistence length of semiflexible polymers.

While the universality of polymer chains in good solvents is well known through scaling exponents, such as the mean square displacement exponent [Eq. (\ref{nuexp})], several universal aspects of polymer shape continue to draw significant research attention.
Our primary goal will be to analyze such    shape properties of semi-flexible polymer chains. The spatial shape attained by a typical polymer conformation in a solvent is of great importance, as it determines the mobility and folding kinetics \cite{Dima2004}, as well as the catalytic activity \cite{Plaxco1998}, of biopolymers. The hydrodynamics of polymer fluids is also strongly affected by the size and shape properties of individual macromolecules \cite{Torre2001}.
To characterize the effective shape of polymers in $d$-dimensional space, it is standard to use quantities such as the average asphericity $\langle A_d\rangle$ and prolateness $\langle S\rangle$; here and below, $\langle\ldots\rangle$ denotes an average over an ensemble of all possible polymer configurations \cite{Aronovitz1986,Rudnick1986}. The asphericity parameter is strictly constrained to the interval $
	0\leq\langle A_d\rangle\leq1$.
A value of zero represents a perfectly symmetric spherical conformation, whereas the upper bound of unity characterizes a fully extended, rigid rod-like configuration.
The quantity $\langle S\rangle$ is defined only for $d=3$. It takes positive values for prolate, ellipsoid-like configurations and negative values for oblate shapes, and is generally bounded by
$
	-\frac{1}{4}<\langle S\rangle<2. $
Rigorous mathematical definitions of these quantities will be given below.

The size and shape parameters of a wide range of flexible polymer macromolecules with various topologies have been analyzed both analytically and numerically \cite{Kalyuzhnyi2019,Ma1994,Paturej2016,Nakamura,Theodorakis2012,Blavatska2015}. This includes complex scale-free polymer networks in which the functionalities, or degrees, of the nodes obey a power-law decaying distribution \cite{Blavatska2025}.
The conformational shape properties of linear and circular semiflexible macromolecules have been extensively analyzed in Refs.~\cite{Alim2007,Chen2020,Fonari2013,Liu2017}. These works demonstrate that the shape parameters $A$ and $S$ change continuously as functions of the persistence length, capturing the structural transition from a disordered flexible coil to a fully extended rigid rod.

L\'evy-based descriptions of polymers have previously been considered
for stiff chains with self-avoidance \cite{Moon1990} and for non-Gaussian
chains with $\alpha$-stable statistics \cite{MajkaGora2015}. Here, we focus on the influence of the L\'evy exponent on global conformational characteristics such as radius of gyration, asphericity, and shape anisotropy. These quantities provide important information about the spatial organization of macromolecular structures and serve as sensitive probes of the underlying stochastic dynamics.
In the present work, we compare polymer conformations generated by persistent random walks and  L\'evy walks in three dimensions, neglecting excluded-volume interactions. 
We determine the radius of gyration $R_{{\rm g}}$ and its effective scaling exponent $\nu$, together with the ensemble-averaged asphericity $\langle A_3\rangle$ and prolateness $\langle S\rangle$. We also analyze the probability distributions of $R_{{\rm g}}$, $A_3$, and $S$. 
By varying the persistence probability $p$ in the first model and the exponent $\sigma$ and upper segment-length cutoff $\ell_{\max}$ in the second, we examine how the segment-length distribution affects these conformational characteristics.
The rest of the paper is organized as follows. Section~\ref{sec:shape-definitions} defines the shape parameters, and Section~\ref{sec:analytical-background} compares the segment-length distributions of the two models. The simulation procedure and  discussion of results obtained are presented in Section~\ref{sec:method}. We follow with the conclusions in Section~\ref{sec:conclusions}.

\section{Polymer shape parameters} \label{sec:shape-definitions}

The spatial shape characteristics  of any polymer macromolecule, represented by a sequence of $N$ monomers within a $d$-dimensional space, can be effectively described \cite{Solc1971} using the gyration tensor $\mathbf{Q}$ with matrix elements defined as follows:
\begin{equation}
Q_{ij}=\frac{1}{N}\sum_{n=1}^N(x_n^i-{x^i_{{\rm CM}}})(x_n^j-{x^j_{{\rm CM}}}),\,\,\,\,\,\,i,j=1,\ldots,d.
\label{mom}
\end{equation}
In this expression, $x_{n}^{i}$  denotes the $i$th Cartesian component of the position vector $\vec{R}_{n} $ of the $n$th monomer, and ${x^i_{{\rm CM}}}=\sum_{n=1}^N x_n^i/N$ represents the $i$th component of the center-of-mass vector $\vec{R}_{{\rm CM}}$. 
The trace of the gyration tensor $\mathbf{Q}$ yields the squared radius of gyration $R_{{\rm g}}^{2}$: 
\begin{equation} 	
R_{{\rm g}}^2 ={\rm Tr}\, {\bf{Q}} =  \sum_{i=1}^d Q_{ii} =\frac{1}{N}\sum_{n=1}^N (\vec{R}_n-{\vec{R}_{{\rm CM}}})^2, \label{rg1}	 
\end{equation}
which serves as a direct measure of the spatial distribution of monomers relative to the center of mass, characterizing the entire chain. A conformation may have a small end-to-end distance while retaining a large radius of gyration.

The anisotropy and shape asymmetry of the polymer can be evaluated based on the eigenvalues $\lambda _{i}$  of the gyration tensor (\ref{mom}). For perfectly isotropic, spherical shapes, all eigenvalues $\lambda _{i}$ degenerate to identical values. A completely elongated, rod-like structure is characterized by a single non-zero eigenvalue, while all others vanish. By defining the arithmetic mean of the eigenvalues as 
${\overline{\lambda}}\equiv {\rm Tr}\, {\bf{Q}}/d$, the asphericity $A_{d}$ is defined as \cite{Aronovitz1986}: 
\begin{equation} 	
A_d=\frac{1}{d(d-1)}\sum_{i=1}^d\frac{(\lambda_i-{\overline{\lambda}})^2}{{\overline{\lambda}^2}}. \label{asfer1} 
\end{equation}
This structural parameter is dependent on the spatial dimension $d$ even within the framework of the ideal Gaussian chain approximation. To evaluate properties that are directly observable in physical experiments, the ensemble average $\langle A_d \rangle$ over all available conformational states is to be performed.  However, exact analytical calculations face  difficulties while treating averaging (\ref{asfer1}) directly.  An alternative mathematical quantity, $\hat{A_{d}}$, was introduced \cite{Aronovitz1986,Rudnick1986,Blavatska2015}: 
\begin{equation} 	
\hat{A_d}=\frac{1}{d(d-1)}\sum_{i=1}^d\frac{\langle (\lambda_i-{\overline{\lambda}})^2 \rangle}{\langle {\overline{\lambda}^2} \rangle}, \label{asfer2} 
\end{equation}
where the numerator and the denominator are averaged independently. The ``true" ensemble-averaged asphericity is typically extracted from computer simulations:
\begin{equation}
\langle A_d \rangle=\left\langle \frac{1}{d(d-1)}\sum_{i=1}^d\frac{(\lambda_i-{\overline{\lambda}})^2}{{\overline{\lambda}^2}} \right\rangle \label{asfer}.
\end{equation}

In what follows below, we will address the three-dimensional case, $d=3$.
The prolateness in three dimensions is defined by
$S=\prod_{i=1}^{3}(\lambda_i-\overline{\lambda})/\overline{\lambda}^{3}$.
It is positive for prolate conformations and negative for oblate ones.
Both $A_3$ and $S$ are evaluated for each configuration before taking the ensemble average.
Correspondingly, numerical calculations of the ``true" averaged asphericity (Eq. (\ref{asfer})) and prolateness give $ \langle A_{3} \rangle_{{\rm RW}}{=}0.394274$ \cite{Wei1997a}, $\langle S \rangle_{{\rm RW}} = 0.47495$ \cite{Jagodzinski1992}.

\section{Analytical background} \label{sec:analytical-background}
We consider two complementary models of semiflexible polymer chains: persistent random walks (PRW) and L\'evy walks (LW). Both generate sequences of straight segments, but differ in their segment-length statistics: the distribution for PRW constitutes the sequence of independent trials whereas it follows a power law for LW. We first examine these distributions and their moments, which provide a basis for comparing the conformations generated by the two models.

{\it Persistent random walk (PRW).}
 In a process of growing the trajectory, the probability to perform the next step in the same direction as the previous one is $p$, whereas the probability of changing direction is $1 - p$. Thus, the probability for a straight segment of length $\ell$ (consisting of $\ell$ steps) to appear reads:
\begin{equation}
P_{{\rm PRW}}(\ell)=p^{\ell-1}(1-p), \quad 0 < p <1. \label{Ppers}
\end{equation}

The average length of a straight segment (in a case of infinitely long chain $N\to \infty$) is given by:
\begin{equation} 
\langle \ell\rangle_{{\rm PRW}} =(1-p)\sum _{\ell=1}^{\infty }\ell\cdot p^{\ell-1}=(1-p)\,\,\frac{{\rm d}}{{\rm d}p}\sum _{\ell=1}^{\infty }p^\ell=\frac{1}{1-p}.\label{kpers}
\end{equation}
In the semiflexible polymer interpretation, $\langle \ell\rangle_{{\rm PRW}}$ can be viewed as persistence length $\ell_{\rm p}$, the characteristic length scale of local rigidity of polymer chain.
Moments of $\ell$ are given in a form:
\begin{equation}
    \langle \ell^n\rangle_{\rm PRW} = (1-p)\sum_{\ell=1}^{\infty}\ell^n p^{\ell - 1},
\end{equation}
and are finite at any $p<1$. In fact, they can be calculated using the Stirling numbers of the second kind \cite{abramowitz1972stirling}:
\begin{equation}
    \langle \ell^n\rangle_{\rm PRW} = \sum_{j=0}^{n}j!S(n+1,j+1) \left( \frac{p}{1-p} \right)^j ,
\end{equation}
where $ S(n,j) = \frac1{j!} \sum_{k=0}^j (-1)^{j-k}\binom{j}{k}k^n$ with $\binom{n}{j}$ being a binomial coefficient.

In particular, the second moment can be calculated as
\begin{equation}
    \langle \ell^2\rangle_{{\rm PRW}}=\frac{(1+p)}{(1-p)^2}. \label{varPers}
\end{equation}

\begin{figure}[h!]
    \centering
    \includegraphics[width=\linewidth]{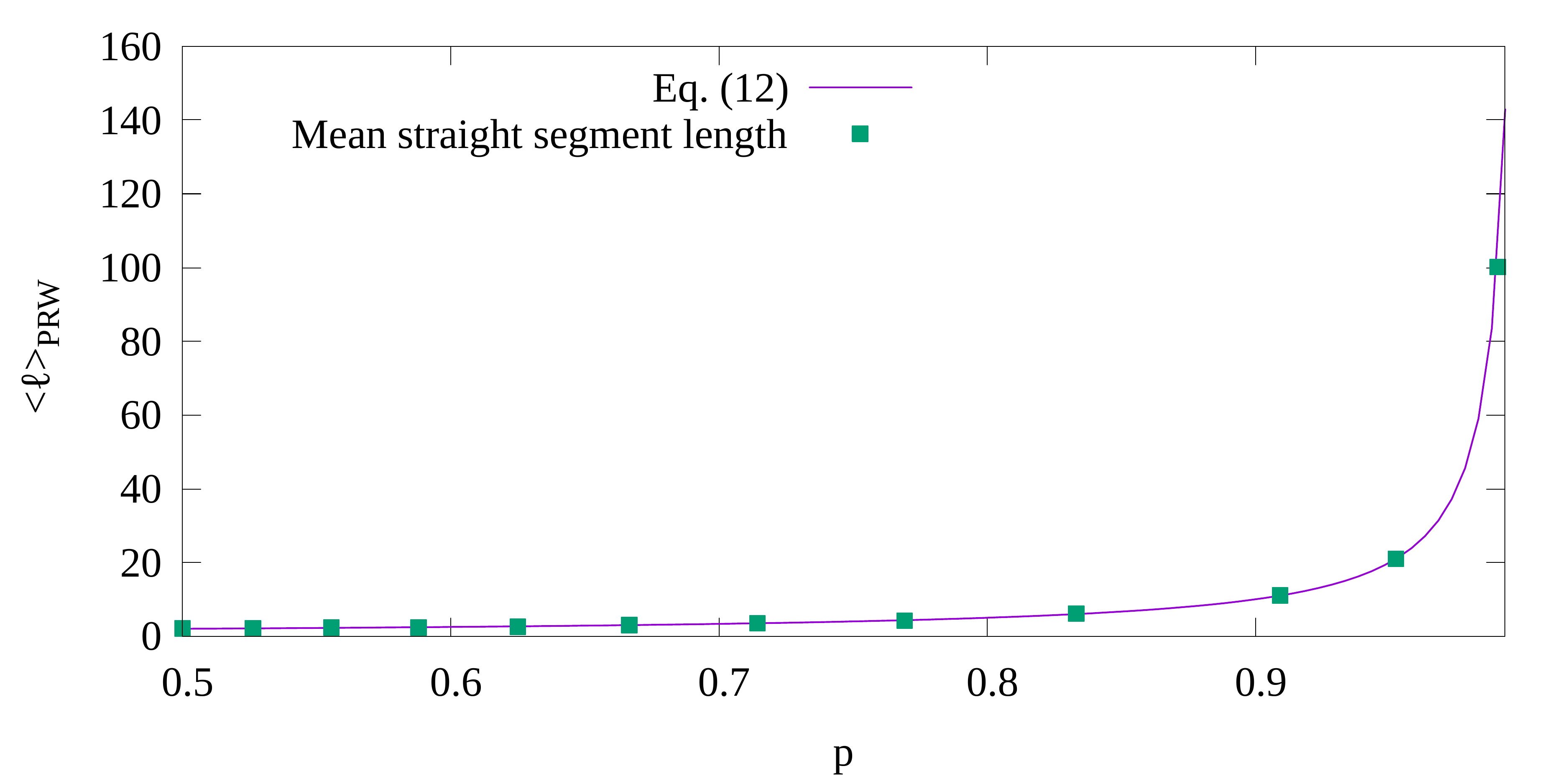}
    \caption{Averaged persistence length $\langle \ell \rangle$ of PRW model as function of  $p$.}
    \label{fig:persistent-kmean}
\end{figure}

With growing $p$, $\langle \ell\rangle_{{\rm PRW}}$ is increasing as expected (see Fig. \ref{fig:persistent-kmean}), and the polymer chain extends towards the perfectly rigid rod state at $p=1$, where $\langle \ell\rangle_{{\rm PRW}} \to \infty$ for an infinite chain. 
At $p<1$, the trajectory still forms a random coil on a large scale, however, the effective step size is lengthened. As we can see from Fig.~\ref{fig:persistent-short},  trajectories of the same length become more extended and elongated in space  as we increase the parameter $p$. The nonlinear
increase of $\langle\ell\rangle_{\rm PRW}$ becomes particularly pronounced as
$p\to1$, providing a direct measure of the persistence introduced by the model.
The divergence in Eq.~(\ref{kpers}) refers to an infinitely long trajectory,
in a finite chain, the segment length is bounded by the chain length.

\begin{figure}[h!]
    \centering
    \includegraphics[width=\linewidth]{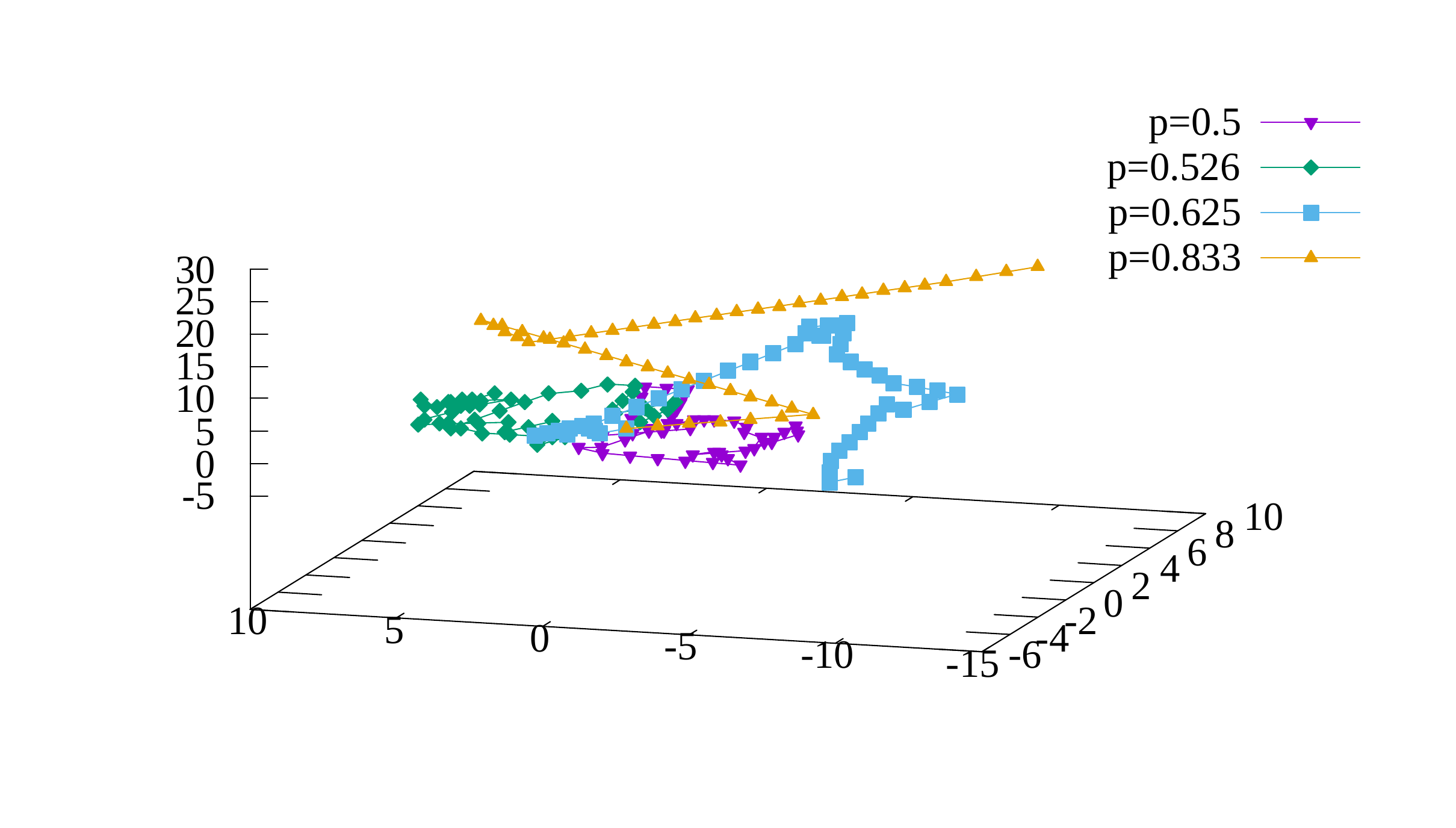}
     \includegraphics[width=\linewidth]{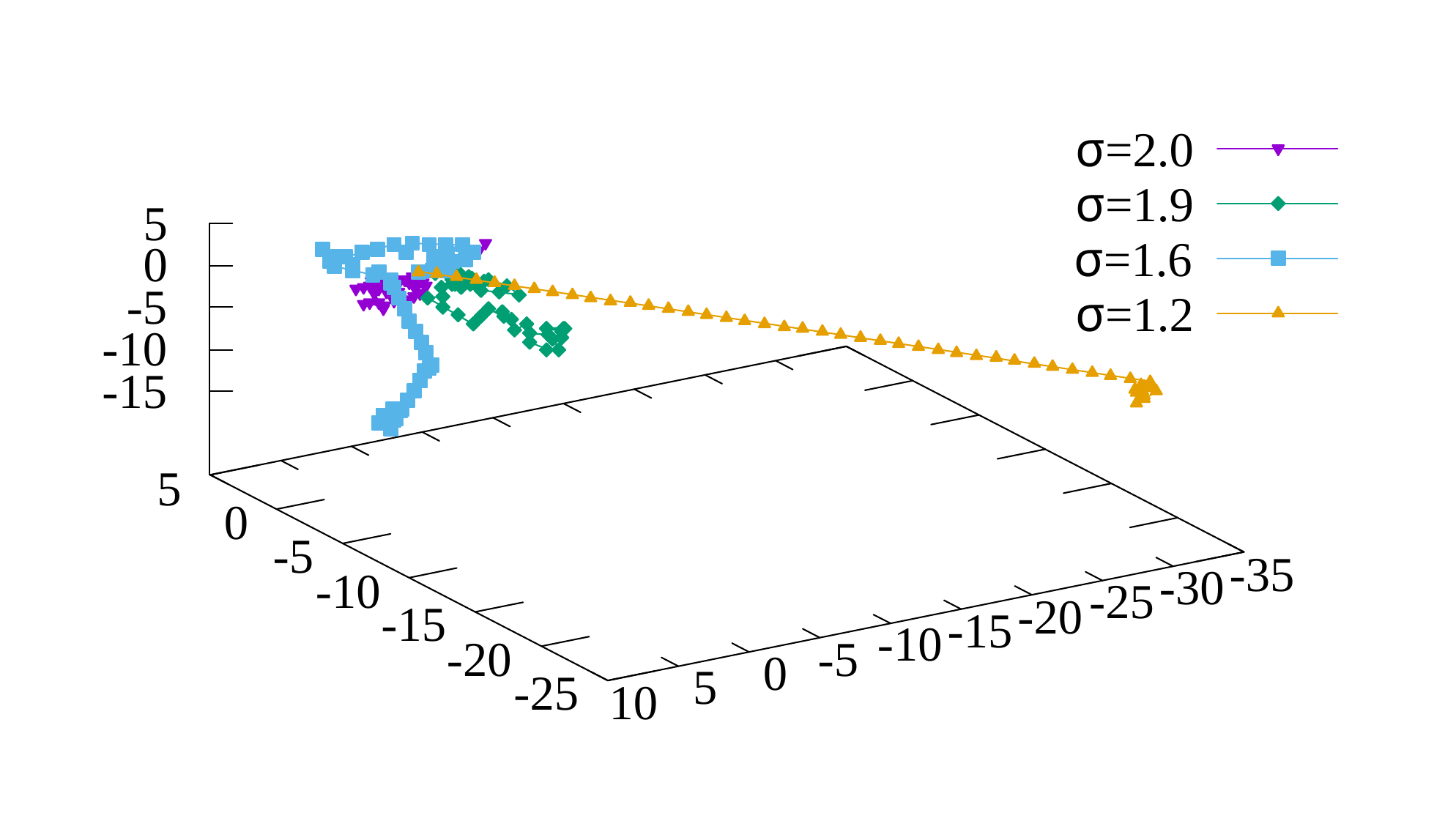}
    \caption{Snapshots of the first 50 steps generated by the two semiflexible polymer models analyzed in this paper: 
the PRW model (upper panel) and the LW model (lower panel). In the PRW model, parameter $p$ defines the persistence probability. 
In the LW model, segment directions are uniformly distributed, and straight-segment lengths $\ell$ follow a power-law 
distribution with decay exponent $\sigma$ [Eq. \ref{PLevy}]. The parameters are related by $p = 1/\sigma$ [Eq. (\ref{kser})], 
ensuring $\langle \ell \rangle_{\rm PRW} = \langle \ell \rangle_{\rm LW}$. Colors denote four different pairs 
of matching parameters $(p, \sigma)$.}
    \label{fig:persistent-short}
\end{figure}

{\it  L\'evy walks (LW)}. The normalized probability density function of the straight segment length $\ell$ at large distances is given by Eq. (\ref{PLevy}).
The average value of segment length is evaluated as:
\begin{equation}
\langle \ell\rangle_{{\rm LW}} = \int _{\ell=1}^{\infty }\ell \cdot \frac{\sigma}{\ell^{1+\sigma }}\,d\ell=\frac{\sigma }{\sigma -1}. \label{kLevy}
\end{equation}
The mean square is given by
\begin{equation}
\langle \ell^2\rangle_{{\rm LW}} =\frac{\sigma }{\sigma -2} \label{varLevy}
\end{equation}
 and is divergent at $\sigma\leq 2$.

We can roughly relate the average segment lengths ($\langle \ell \rangle_{\rm PRW} = \langle \ell \rangle_{\rm LW}$) as given by Eqs. (\ref{kpers}) and (\ref{kLevy}) by:
\begin{equation}
\frac{1}{1-p}=\frac{\sigma}{\sigma-1 }\rightarrow \sigma=\frac{1}{p },
\label{kser}
\end{equation}
claiming that by appropriate choice of parameters $p$ and $\sigma$ in these two models, we can observe the same value  of $\langle \ell\rangle$ in both of them.  Fig. \ref{fig:persistent-short} presents snapshots of several LW trajectories related to PRW trajectories shown in Fig. \ref{fig:persistent-short} (with the values of $\sigma$ being the inverse of $p$ values there, correspondingly). 
Note that it does not imply the fundamental analogy between models (\ref{Ppers}) and (\ref{PLevy}).  Even if we tune $p$ and $\sigma $ in such a way that their first moments $\langle \ell \rangle$ match exactly, the global geometry of the constructed polymer configurations will differ due to fluctuations and higher-order moments.  In particular, in  the L\'evy model, the distribution possesses a ``heavy tail'':  even when $\langle \ell\rangle_{{\rm LW}}$  is finite (at $\sigma > 1$), the second moment  diverges for $1 <   \sigma \le 2$. 

\begin{figure}[ht]
    \centering
    \includegraphics[width=\linewidth]{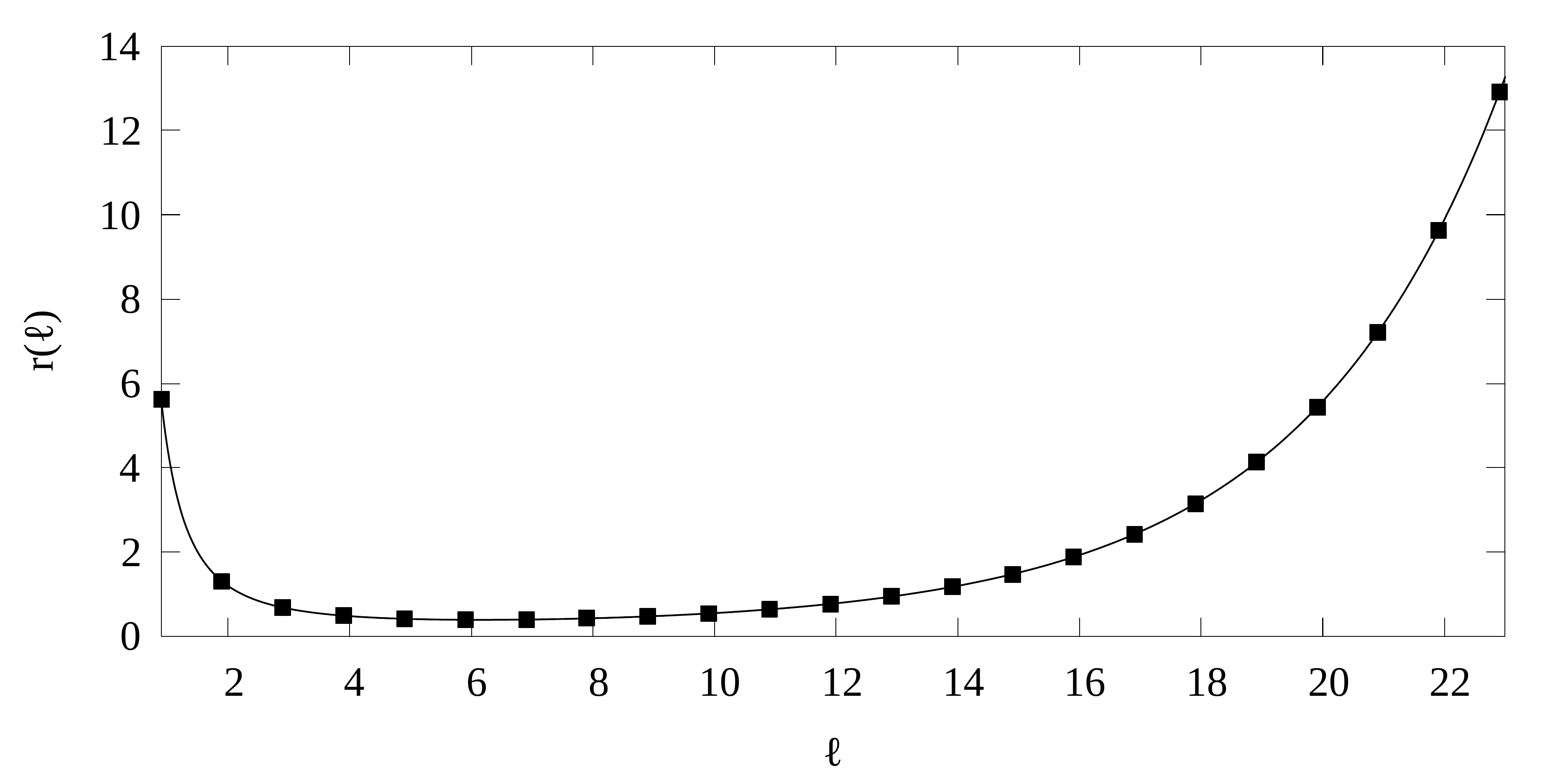}
    \caption{ The relative probability $r(\ell)$ as given by Eq. \ref{rat} at fixed $\sigma=1.5$}
    \label{RatioP}
\end{figure}

Let us introduce the ratio $r(\ell)$ of probabilities (\ref{PLevy}) and (\ref{Ppers}) taken under the condition $p = 1/\sigma$: 
\begin{equation}
r(\ell)=\frac{P_{{\rm LW}}(\ell)}{P_{{\rm PRW}}(\ell)}=\frac{\sigma^{\ell+1}}{(\sigma-1)\ell^{1+\sigma}}.
\label{rat}
\end{equation}
 and presented at a fixed value of  $\sigma=1.5$  in Fig. \ref{RatioP}. 
 At small $\ell$, the power-law decay of the numerator is balanced by the denominator, making the probabilities comparable. However, in the asymptotic regime of large $\ell$ ($\ell \gg 1$), the exponential growth of the numerator $\sigma^{\ell+1}$ plays the  dominant role. This effect is caused by the non-trivial probability of  giant ``jumps''  to occur in L\'evy flight trajectory that are exponentially suppressed within the PRW model.
}

In what follows below, we use numerical simulations to compare the global conformations generated within the two above described models, Eqs.~\ref{PLevy} and~\ref{Ppers}, correspondingly. We generate ensembles of three-dimensional chains and evaluate the size and shape observables defined in Sec.~\ref{sec:shape-definitions}.

\section{Results and discussions} \label{sec:method}
\subsection{The method}
Both models were simulated in three-dimensional continuous space. For the L\'evy-walk model, the length $\ell$ of each straight segment was sampled from the truncated probability distribution Eq. (\ref{PLevy}). In such distribution the upper cutoff $\ell_{\max}$ specifies the largest allowed length of an individual straight segment. The density is zero outside the interval $[1,\ell_{\max}]$. Truncation preserves the power-law dependence within the allowed range but excludes segments longer than $\ell_{\max}$, making all segment-length moments finite. Varying $\ell_{\max}$ allows us to assess how restricting the heavy tail affects the size and shape of the resulting polymer conformations. The optimal truncation is discussed in what follows.

The direction of each segment was chosen independently and isotropically by generating a random unit vector in spherical coordinates. Specifically, the angle $\phi$ was sampled uniformly from $[0, 2\pi]$, while $\cos \theta$ was sampled uniformly from $[-1,1]$. This procedure ensures a uniform distribution of directions over the unit sphere.
In the PRW model, each unit step retained the direction of the preceding step with probability $p$. With probability $(1-p)$, a new isotropically distributed direction was generated using the same spherical-coordinate procedure as in the L\'evy-walk model. Consequently, each trajectory consisted of straight segments whose lengths followed the distribution given by Eq. (\ref{Ppers}).
For the main data set, we generated ensembles of $M=10^5$ independent configurations, each containing $N=10^5$ steps (monomers). For the LW model, the upper cutoff $\ell_{\max}$ ranged from $10^2$ to $10^7$.
The observables were calculated separately for each
configuration at fixed $N$ and subsequently averaged
over the ensemble.

Additional calculations at $\ell_{\max}=10^{10}$ were performed with substantially lower precision.
These results are included in Fig.~\ref{fig:nus-as-kmax} as a qualitative reference for the trend at large cutoffs.

Using this procedure, we first examine PRW model as a reference, and then analyze how the size and shape of LW conformations depend on the distribution parameters.

\subsection{Persistent random walks}
\begin{figure}[ht]
    \centering
    \includegraphics[width=0.95\linewidth]{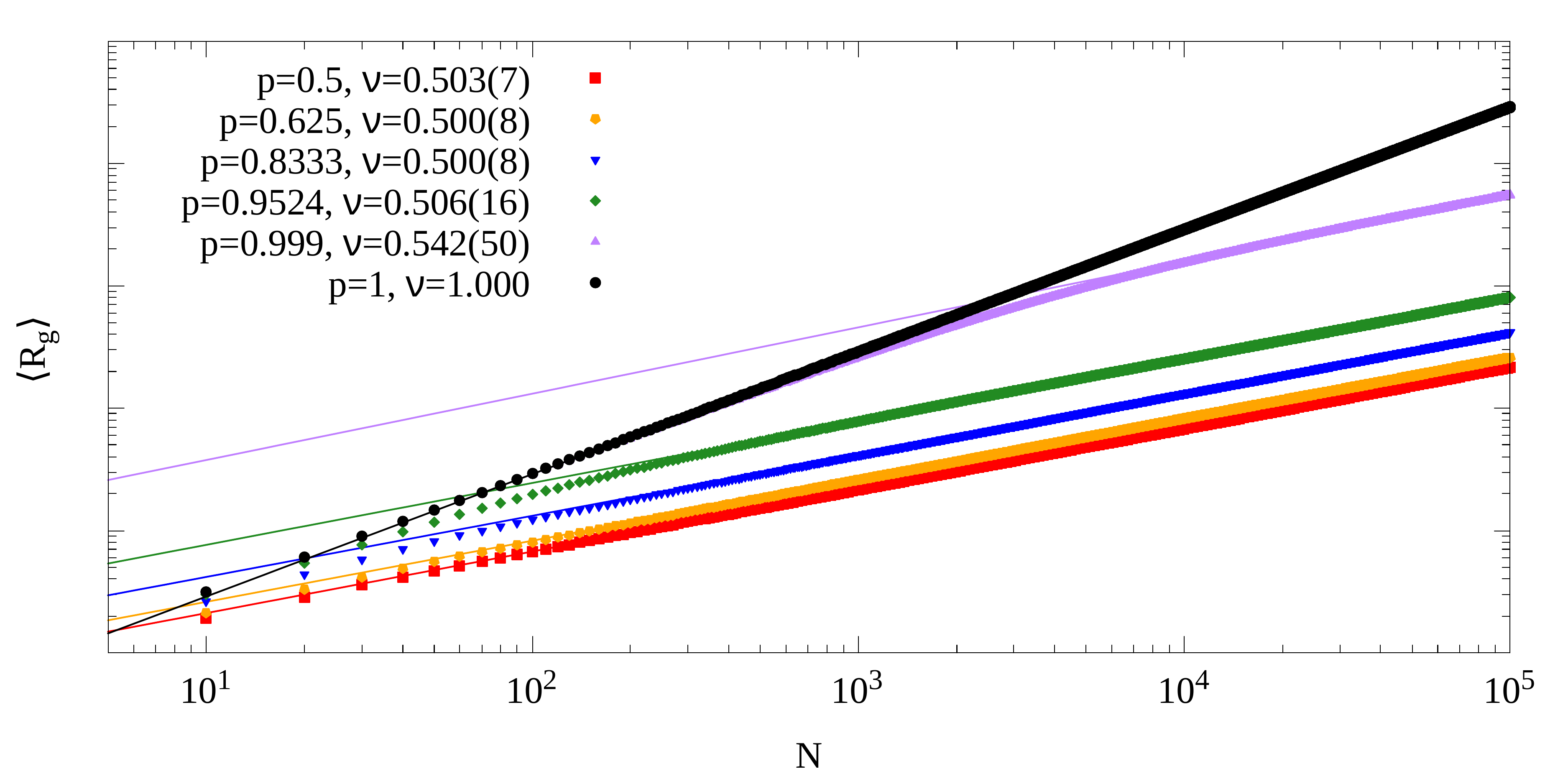}
    \includegraphics[width=\linewidth]{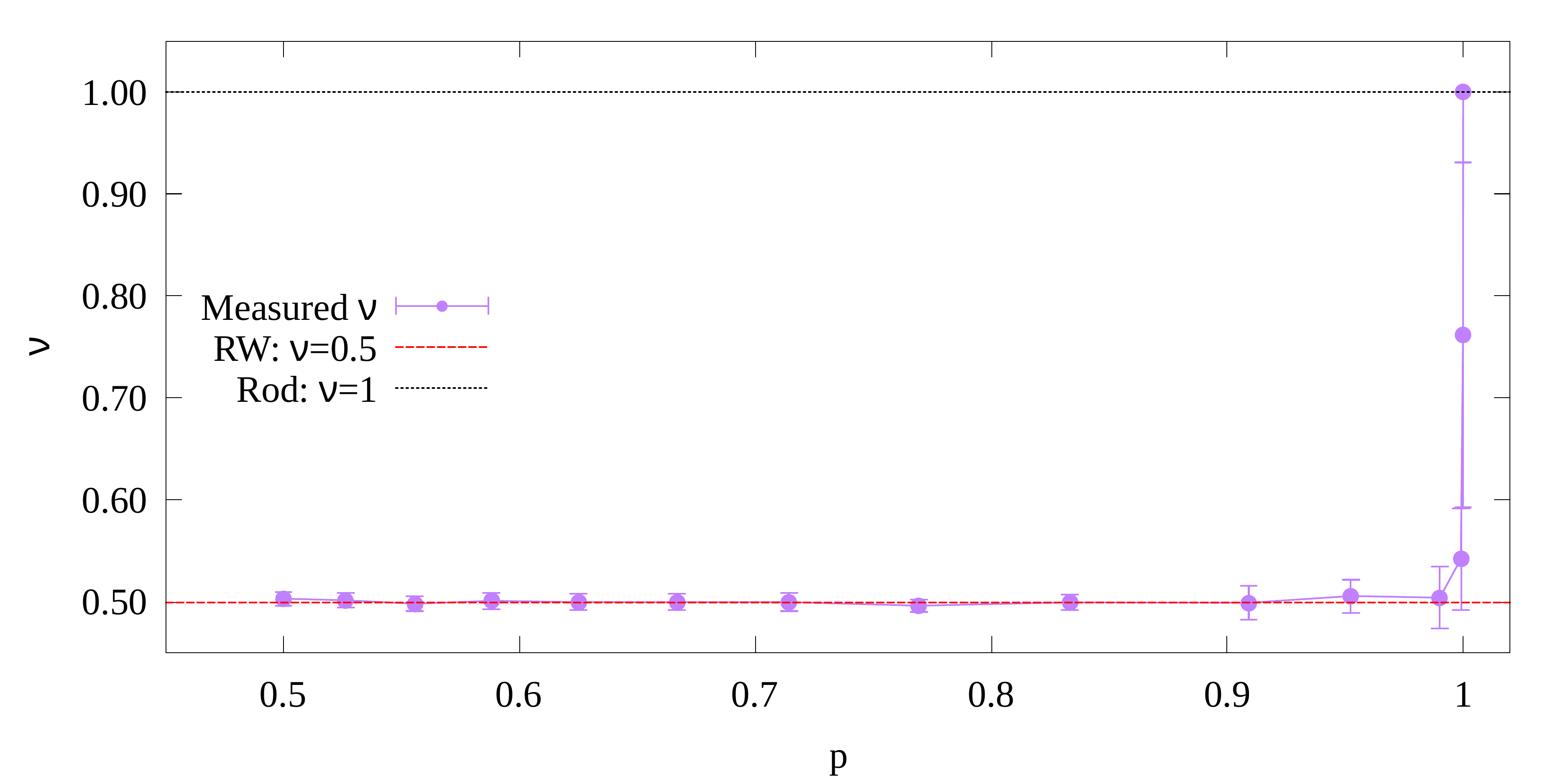}
    \caption{Finite-size scaling of the gyration radius (upper panel) and corresponding exponents $\nu$ (lower panel) for different values of persistence probability $p$. The values of $p$ were chosen to match via relation (\ref{kser}) values of 
$\sigma = 2$ , $1.6$, $1.2$, $1.05$, $1.00$ correspondingly. FSS shows parallel lines (ergo the same power-law exponents) up until values very close to 1. Even for very large $p=0.999$, there is a tendency towards scaling with $\nu=1/2$ as the chain length grows.}
    \label{fig:persistent-results}
\end{figure}

\begin{figure}[h]
    \centering
    \includegraphics[width=\linewidth]{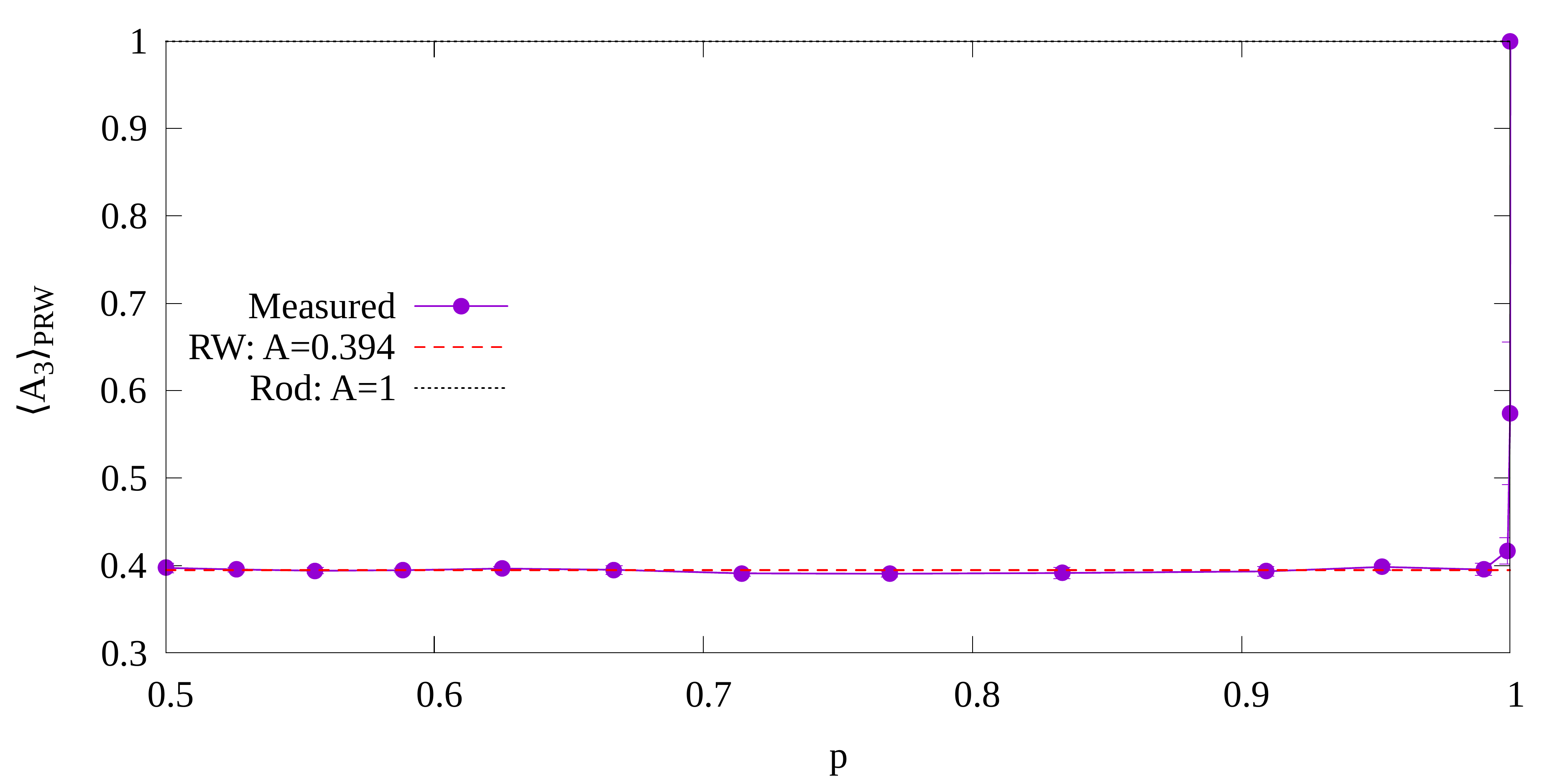}
    \includegraphics[width=\linewidth]{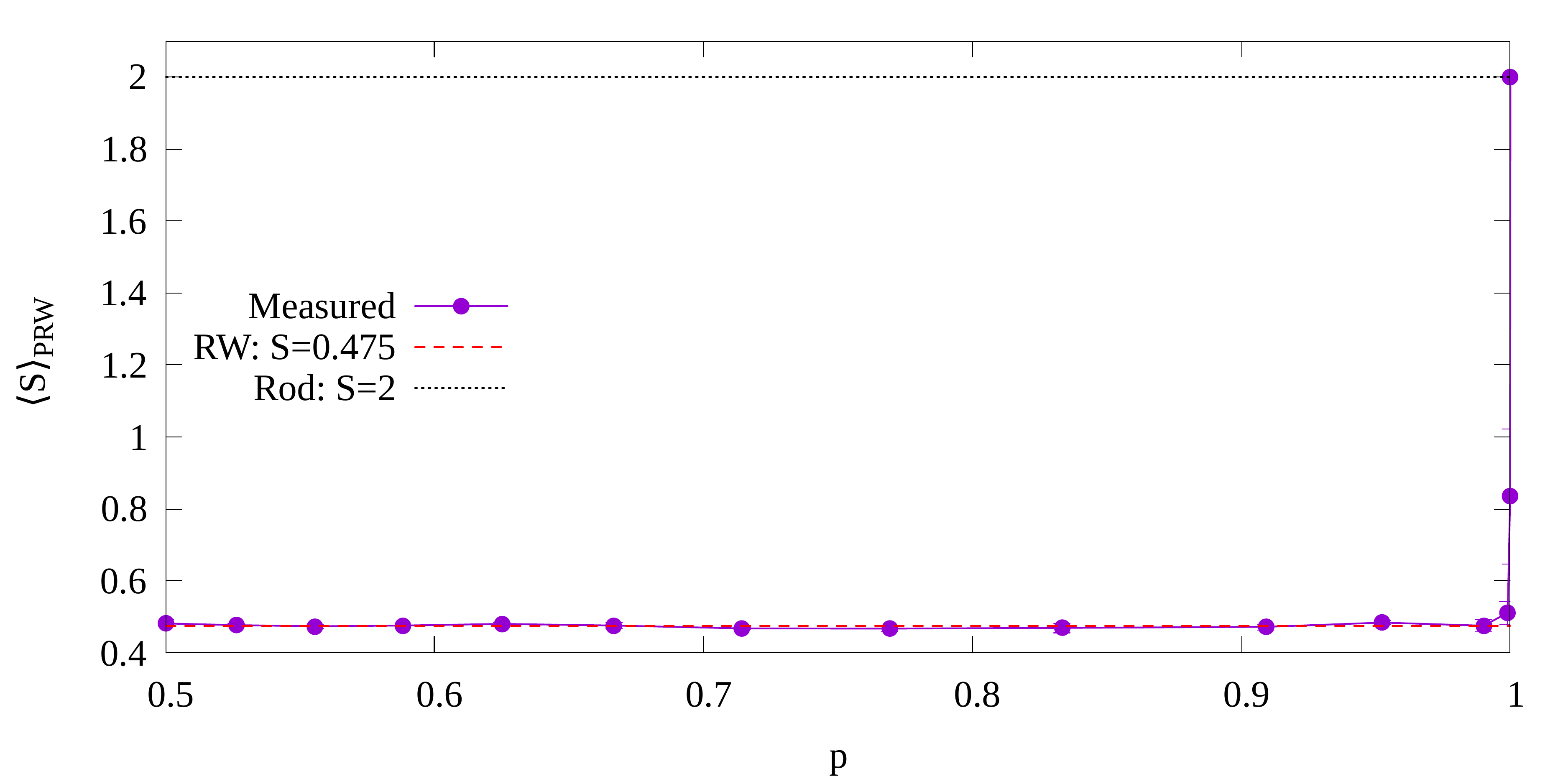}
    \caption{Asphericity $\langle A_3 \rangle_{{\rm PRW}}$ (upper panel) and prolateness $\langle S\rangle_{{\rm PRW}}$ (lower panel) of PRW for different values of the persistence probability $p$. Error bars fit within the data points. Note that, on the given scale, the dashed RW line overlaps with the data fit over almost the entire range of $p$.}
    \label{fig:persistent-shapes}
\end{figure}

Results for the scaling exponent $\nu$ and shape characteristics $A$ and $S$ are given in Figs. \ref{fig:persistent-results} and
\ref{fig:persistent-shapes}.

Across most of the investigated range of $p$, the effective size exponent remains close to the RW (Gaussian) value $\nu=1/2$ (see Fig.~\ref{fig:persistent-results} for the finite-size scaling of $R_{{\rm g}}$), and the mean asphericity and prolateness stay near their RW reference values, see Fig.~\ref{fig:persistent-shapes}. This behavior holds even at high persistence probabilities. Significant structural changes appear only very close to $p=1$, where the chain stretches into strongly elongated conformations that approach the rigid-rod limit ($\nu=1$, $\langle A_3\rangle=1$, and $\langle S\rangle=2$ at $p=1$). This trend aligns with a finite-chain crossover governed by the persistence length scale in Eq.~(\ref{kpers}), through the ratio
\begin{equation}
\frac{N}{\langle\ell\rangle_{\rm PRW}}=N(1-p).
\end{equation}
For $N(1-p)\gg1$, the trajectory behaves like an ordinary random walk on large length scales. When $N(1-p)$ drops to order unity or below, direction changes become rare and the chain straightens out. Because of this, the crossover sharpens toward $p=1$ as $N$ increases. At any fixed $p<1$, a sufficiently long chain eventually recovers the Gaussian regime, whereas $p=1$ yields a perfectly straight rod regardless of $N$. The sharp transition near $p=1$ thus traces back to the interplay between chain length and persistence length. Consequently, one concludes that the PRW model does not properly describe the expected behavior of a semi-flexible polymer \cite{Alim2007,Chen2020,Fonari2013,Liu2017}, since for almost all values of $p$ it belongs to the universality class of a random walk. Therefore, we focus on the LW  model in the following.

\subsection{L\'evy walks. Gyration radius and exponent $\nu$}

\begin{figure}
    \centering
    \includegraphics[width=\linewidth]{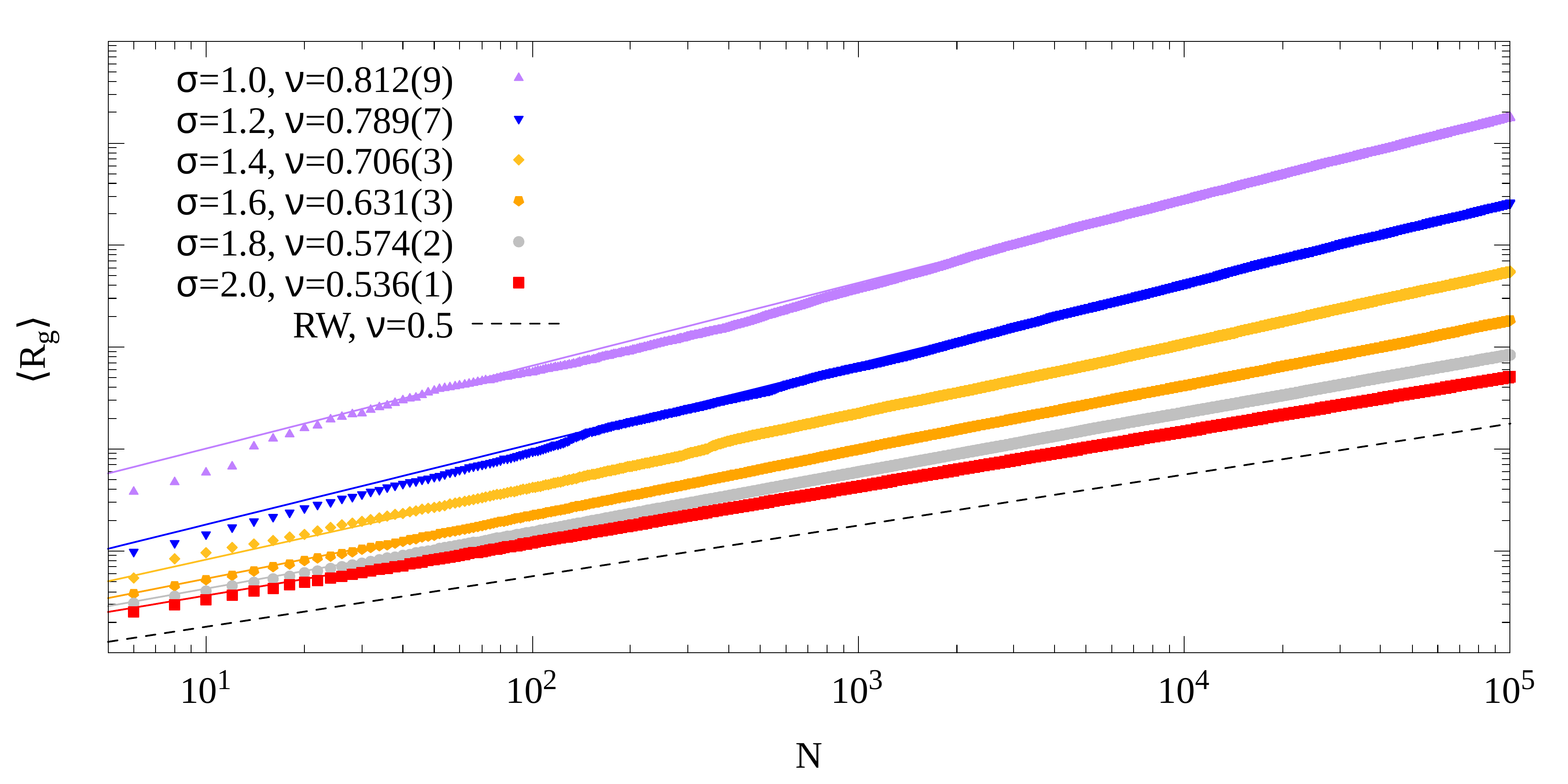}
    \caption{Gyration radius $ \langle R_{{\rm g}}\rangle$ as a function of LW trajectory length $N$  at various  $\sigma$ and fixed $\ell_{\rm max}=10^7$. Since the leading term of $\langle R_{{\rm g}}\rangle (N)$ is power-law, in log-log scale we see straight lines as the walk length grows. }
    \label{fig:r-g-fss}
\end{figure}

{We start with examining the scaling relation (\ref{nuexp}), remembering that analytical prediction for the case of infinitely long L\'evy trajectory is $\nu=1/\sigma$ \cite{Fisher72,Grassberger85}.} 
Figure~\ref{fig:r-g-fss} shows how the gyration radius scales with the chain length at a fixed  upper cutoff value $\ell_{\max}=10^7$. On logarithmic axes, the curves become roughly linear for large $N$, signaling power-law scaling. The variation in their slopes demonstrates that $\sigma$ influences both the magnitude of $R_{{\rm g}}$ and its effective growth exponent. To quantify this behavior, we fit the data to the two-parameter form $\log R_{{\rm g}}(N)=a+\nu \log N$. Across all the shown curves, the fitted exponent grows from $\nu \approx 0.54$ at $\sigma=2.0$ up to $\nu \approx 0.81$ at $\sigma=1.0$. This steeper scaling at smaller $\sigma$ marks a clear departure from the Gaussian baseline of $\nu=1/2$. Note that these exponents describe the accessible chain-length window at fixed cutoff, and determining their true asymptotic limits requires a separate finite-size analysis.

An  appropriate value of cutoff  $\ell_{\rm max}$  can be estimated as follows. Based on  Eq. (\ref{PLevy}), where an infinite upper cutoff is assumed, the probability that the step length $\ell$ will exceed some prescribed value denoted as $\ell_{\rm max}$ is:
\begin{equation}
P(\ell>\ell_{\rm max})= \sigma\int_{\ell=\ell_{\rm max}}^{\infty }\ell^{-1-\sigma }d\ell = \ell_{\rm max}^{-\sigma }.
\end{equation}
For the $N$-step polymer chain, the average number of cases when $\ell$  exceeds  $\ell_{\rm max}$ is thus $N\ell_{{\rm max}}^{-\sigma}$. Only the case when $N \ell_{{\rm max}}^{-\sigma} \ll 1$ means the adequate choice of $\ell_{{\rm max}}$. At fixed $N$,  $\ell_{\rm max}$ should be  large enough to satisfy this condition. Taking into account that we  are modeling an ensemble of $M$ independent trajectories ($M \sim 10^3 - 10^5$), each of the length $N$, the total number of steps generated  is $N_{\rm total} = M \cdot N$. Thus, to observe the correct behavior, the cutoff parameter should satisfy the condition $\ell_{\rm max} \gg (N\cdot M)^{1/\sigma}$.

\begin{figure}
    \centering
    \includegraphics[width=\linewidth]{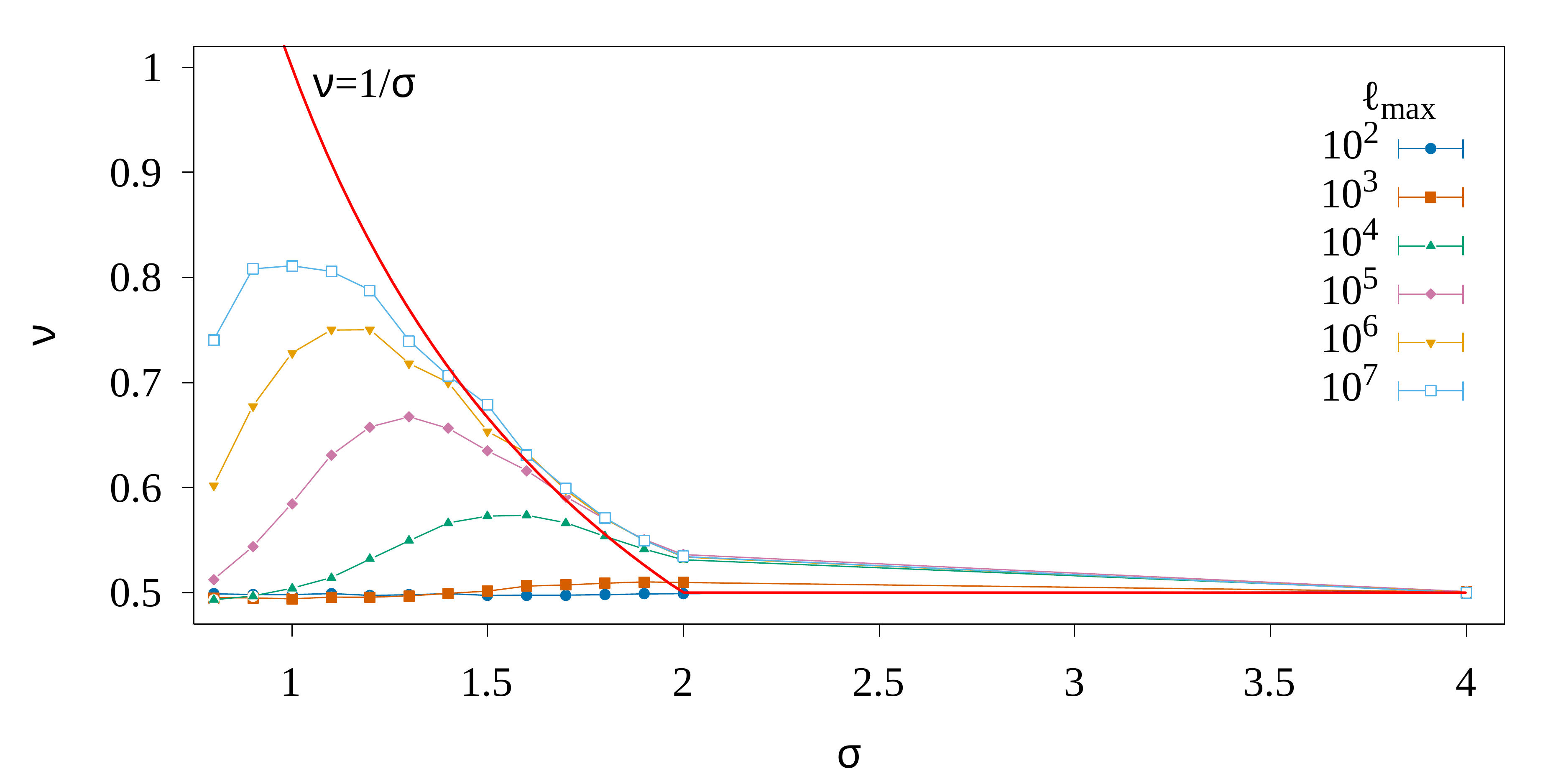}
    \caption{Effective exponent $\nu$ as a function of $\sigma$ for different upper cutoffs $\ell_{\max}=10^2,\ldots,10^7$.  Lines connecting numerical points are eye guides. The solid red curve shows the reference relation $\nu=1/\sigma$ for $1<\sigma<2$, continued by the Gaussian value $\nu=1/2$ for $\sigma\ge2$.}
    \label{fig:nus-compare}
\end{figure}

\begin{figure}
    \centering
    \includegraphics[width=\linewidth]{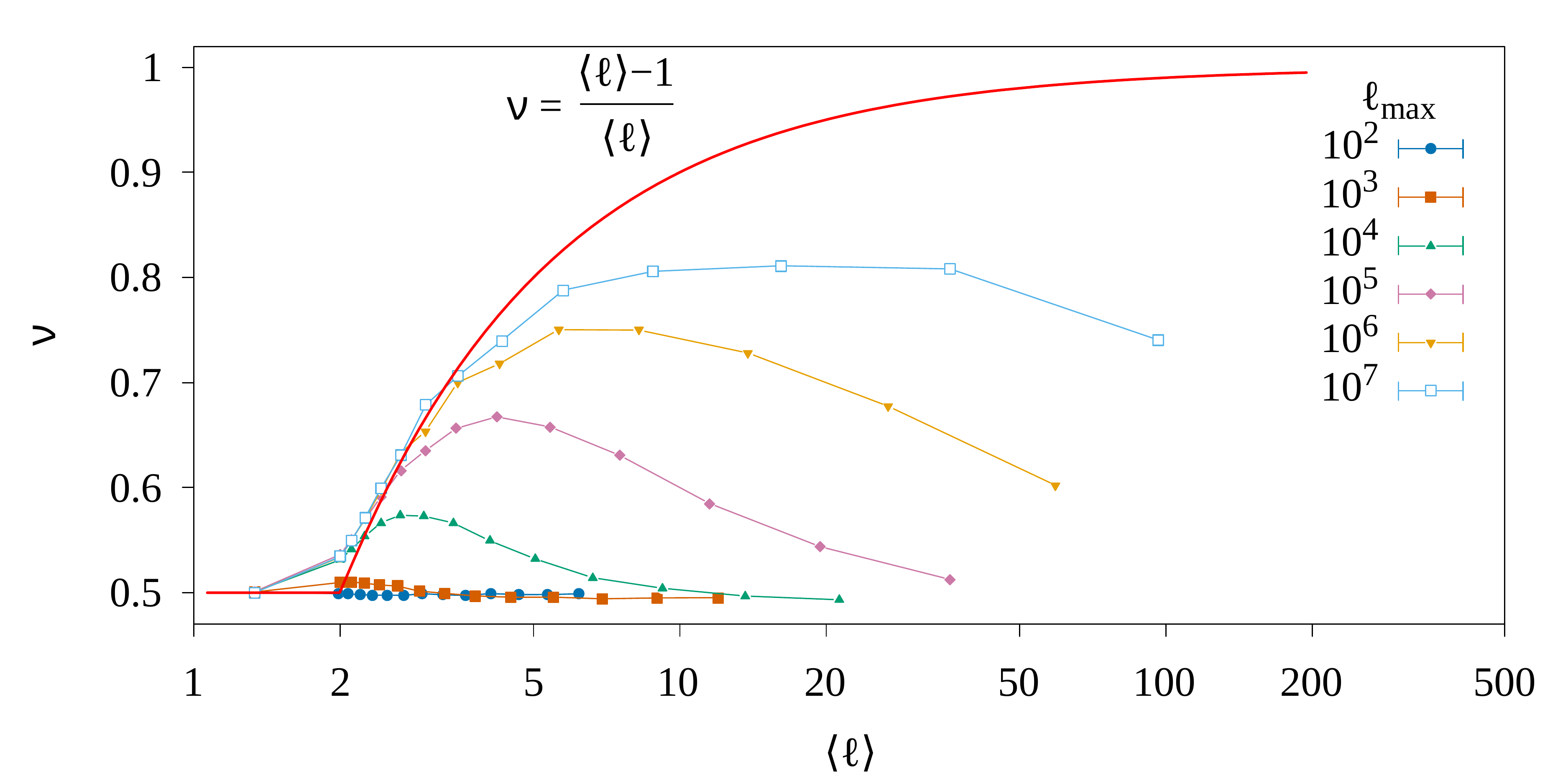}
    \caption{Effective exponent $\nu$ as a function of the mean segment length $\langle\ell\rangle$ (controlled by parameter $\sigma$) for different values of upper cutoffs $\ell_{\max}$. The red line shows the reference relation $\nu=(\langle\ell\rangle-1)/\langle\ell\rangle$ for
$\langle\ell\rangle\ge2$, continued by the Gaussian value $\nu=1/2$ for $1<\langle\ell\rangle<2$.}
    \label{fig:nus-as-mean}
\end{figure}

\begin{figure}
    \centering
    \includegraphics[width=\linewidth]{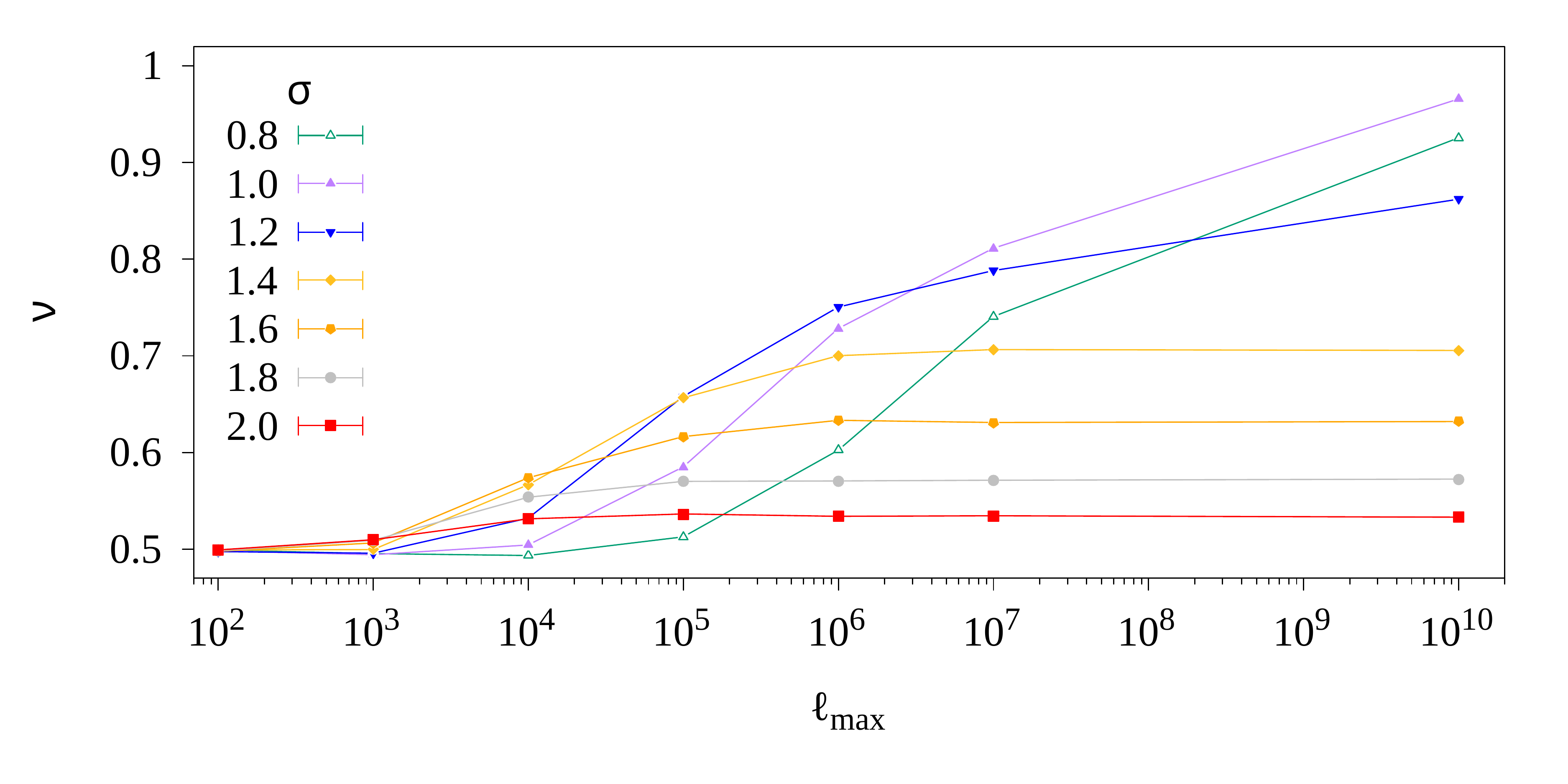}
    \caption{Effective exponent $\nu$ as a function of the upper cutoff $\ell_{\max}$ at variable  $\sigma$. The points at
$\ell_{\max}=10^{10}$ were obtained with  substantially lower precision and are included as a qualitative reference.}
    \label{fig:nus-as-kmax}
\end{figure}

Figure~\ref{fig:nus-compare} (and, consequently, Fig.~\ref{fig:nus-as-mean}) and Fig.~\ref{fig:nus-as-kmax} present two
complementary views of the same fitted exponents: the dependence on
$\sigma$ at fixed $\ell_{\max}$, and the dependence on $\ell_{\max}$ at
fixed $\sigma$.

Figure~\ref{fig:nus-compare} shows the effective exponent $\nu$ as a function of $\sigma$ for various values of upper cutoff in the range $\ell_{\max}=10^2,\ldots,10^7$. For $1<\sigma<2$, the red line represents the reference relation $\nu=1/\sigma$; for $\sigma\ge2$, it shows the Gaussian value $\nu=1/2$.
At the smallest cutoff, the fitted exponents remain close to
the Gaussian reference $\nu=1/2$. Increasing the cutoff makes the
dependence on $\sigma$ more pronounced, indicating that rare long
segments affect the size growth within the simulated length range.
For sufficiently large value of $\ell_{{\rm max}}$, the numerical estimates are merging towards the theoretical reference  for $\sigma\gtrsim1.4$. 
For large $\sigma\geq2$, the results remain close to the Gaussian value
and depend only weakly on the cutoff.
The cutoff dependence becomes more pronounced as $\sigma$ decreases. Even at $\ell_{\max}=10^7$, the effective exponent deviates from the analytical prediction at $\sigma<1.4$. Thus, within the simulated range, the dependence on $\sigma$ is non-monotonic. The revealed discrepancy indicates that for small $\sigma $ the rate of convergence to the asymptotic limit is extremely slow. The theoretical prediction $\nu = 1/\sigma$ is formulated for infinite trajectories without restrictions on the step length. In numerical simulations as well as in experiments, the introduction of any finite cutoff $\ell_{\rm max}$ (even as large as $10^{7}$) creates a strong finite-size effect. The long L\'evy ``jumps", which should determine the superdiffusion nature of the motion at $\sigma < 1.4$, are artificially truncated, which leads to an underestimation of the actual exponent $\nu $.

Figure~\ref{fig:nus-as-mean} presents the effective exponent as a function of the numerically determined mean segment length $\langle\ell\rangle$. For the distribution with infinite upper cutoff, Eq.~(\ref{kLevy}) gives
$\langle\ell\rangle=\sigma/(\sigma-1)$ for $\sigma>1$.
In the range $1<\sigma<2$, rewriting the reference relation $\nu=1/\sigma$ therefore yields
\begin{equation}
    \nu=
    \frac{\langle\ell\rangle-1}
         {\langle\ell\rangle}.
    \label{eq:nu-mean-reference}
\end{equation}
Together with the Gaussian value $\nu=1/2$ for $\sigma\ge2$, this defines the red reference curve.
The numerical curves are merging close to the reference curve near
$\langle\ell\rangle=2$ but deviate as the mean
segment length increases. Thus, within the investigated range, the influence of truncation remains visible in this representation and cannot be neglected.

Figure~\ref{fig:nus-as-kmax} shows the dependence of the effective exponent $\nu$ on the upper cutoff $\ell_{\max}$ for fixed values of $\sigma$.
For $\sigma=1.4,1.6,1.8$, the curves approach nearly constant values at cutoffs of approximately $10^5$--$10^6$. These values are merging the reference relation $\nu=1/\sigma$.
For $\sigma=2.0$, the fitted exponent remains close to
the Gaussian value $\nu=1/2$, albeit slightly above it, throughout the investigated range. For smaller $\sigma$, the cutoff dependence persists over a broader range. In particular, for $\sigma=0.8$ and $1.0$, increasing $\ell_{\max}$ produces a substantial increase in the effective exponent, with no clear plateau within the main range $\ell_{\max}\le10^7$.
The additional results at $\ell_{\max}=10^{10}$ extend
this comparison to larger cutoffs, but only as a general trend due to the reduced precision.
At small cutoffs  the heavy tails of the distribution are completely cut off. The increase in $\ell_{\rm max}$ activates the superdiffusion properties of the trajectories, which leads to a pronounced increase in exponent $\nu $.

\begin{figure}[b!]
    \centering
    \includegraphics[width=\linewidth]{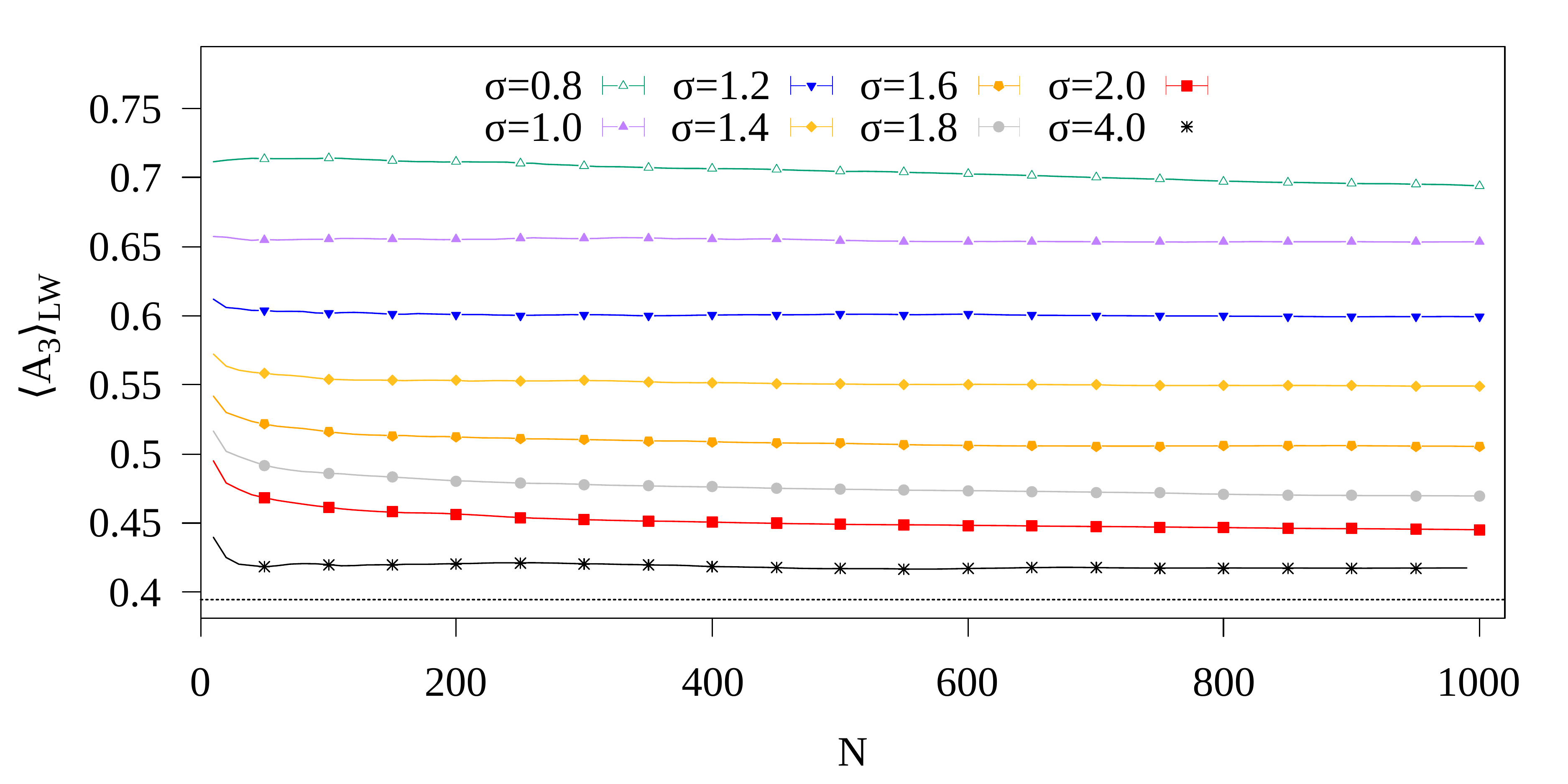}
    \caption{Asphericity $\langle A_3\rangle_{{\rm LW}}$ as a function of the chain length $N$ for different values of $\sigma$ at $\ell_{\max}=10^5$.
The bottom dotted line shows the Gaussian-chain value
$\langle A_3\rangle_{{\rm RW}}=0.394274$ \cite{Wei1997a}.}
    \label{fig:asph}
\end{figure}

This analysis convincingly proves that effective scaling behavior is strongly affected by truncation, especially for small $\sigma$. For distributions with heavier tails, the fitted exponents remain sensitive to $\ell_{\max}$ over a broader range of cutoffs. The cutoffs required to reduce these effects have therefore to be assessed for each $\sigma$ and chain-length range. Determining the asymptotic regime requires a separate analysis of the chain-length dependence.

\subsection{L\'evy walks. Shape characteristics}

\begin{figure}[t!]
    \centering
    \includegraphics[width=\linewidth]{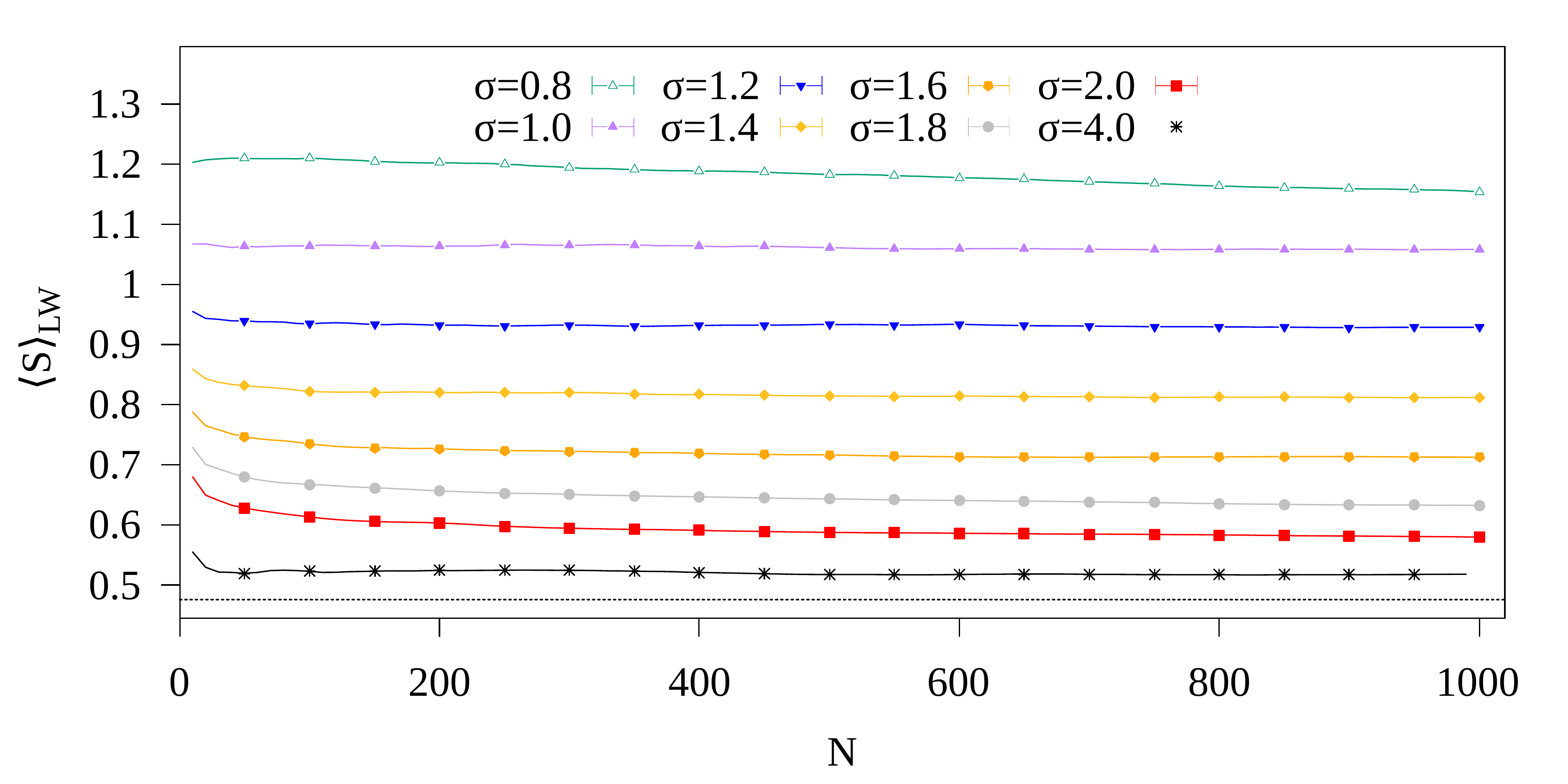}
    \caption{Prolateness $\langle S\rangle_{{\rm LW}}$ as a function of the chain length $N$ for different values of $\sigma$ at $\ell_{\max}=10^5$. Error bars represent standard errors of the mean, the bottom dotted line shows the Gaussian-chain value
$\langle S\rangle_{{\rm RW}}=0.47495$ \cite{Jagodzinski1992}.}
    \label{fig:prol}
\end{figure}

Figures~\ref{fig:asph} and~\ref{fig:prol} show the dependence of the mean asphericity $\langle A_3 \rangle$ and prolateness $\langle S \rangle$ on the chain length $N$ at fixed value of cutoff $\ell_{\max}=10^5$. After a short initial transient, both quantities gradually approach nearly constant values. Thus, although the characteristic size of a trajectory continues to grow with $N$, its  shape becomes effectively independent of the chain length.
The limiting values of both shape parameters depend strongly on the L\'evy exponent $\sigma$. For large $\sigma$, the results approach the corresponding values for a Gaussian chain,
$\langle A_{3}\rangle_ {\rm RW}=0.394274$ \cite{Jagodzinski1992}, and
$\langle S\rangle_{{\rm RW}} = 0.47495$ \cite{Jagodzinski1992}. 
The specific dependence on $\sigma$ also varies with the upper cut-off $\ell_{\rm max}$, as discussed below.

\begin{figure}
    \centering
    \includegraphics[width=\linewidth]{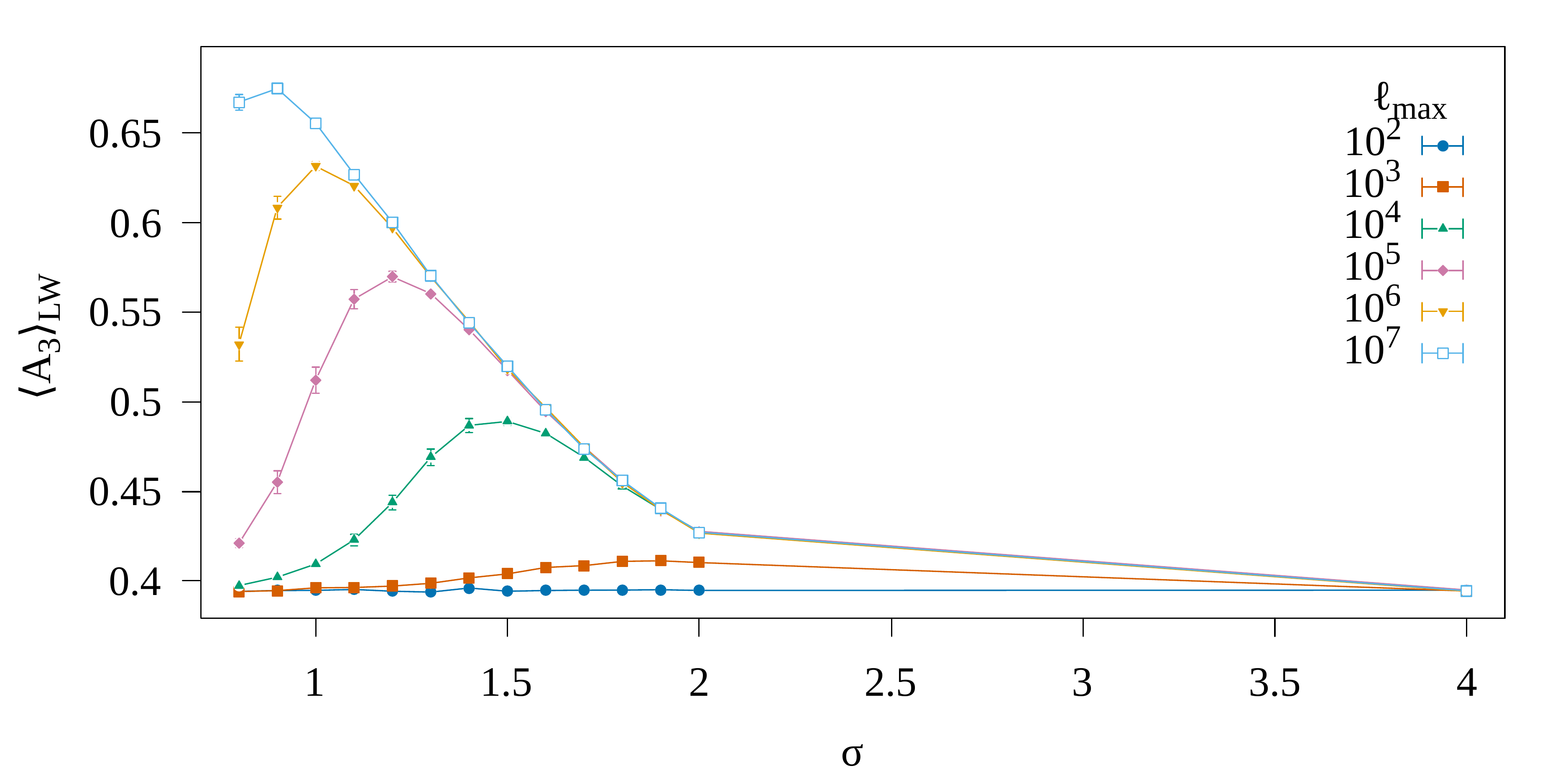}
    \includegraphics[width=\linewidth]{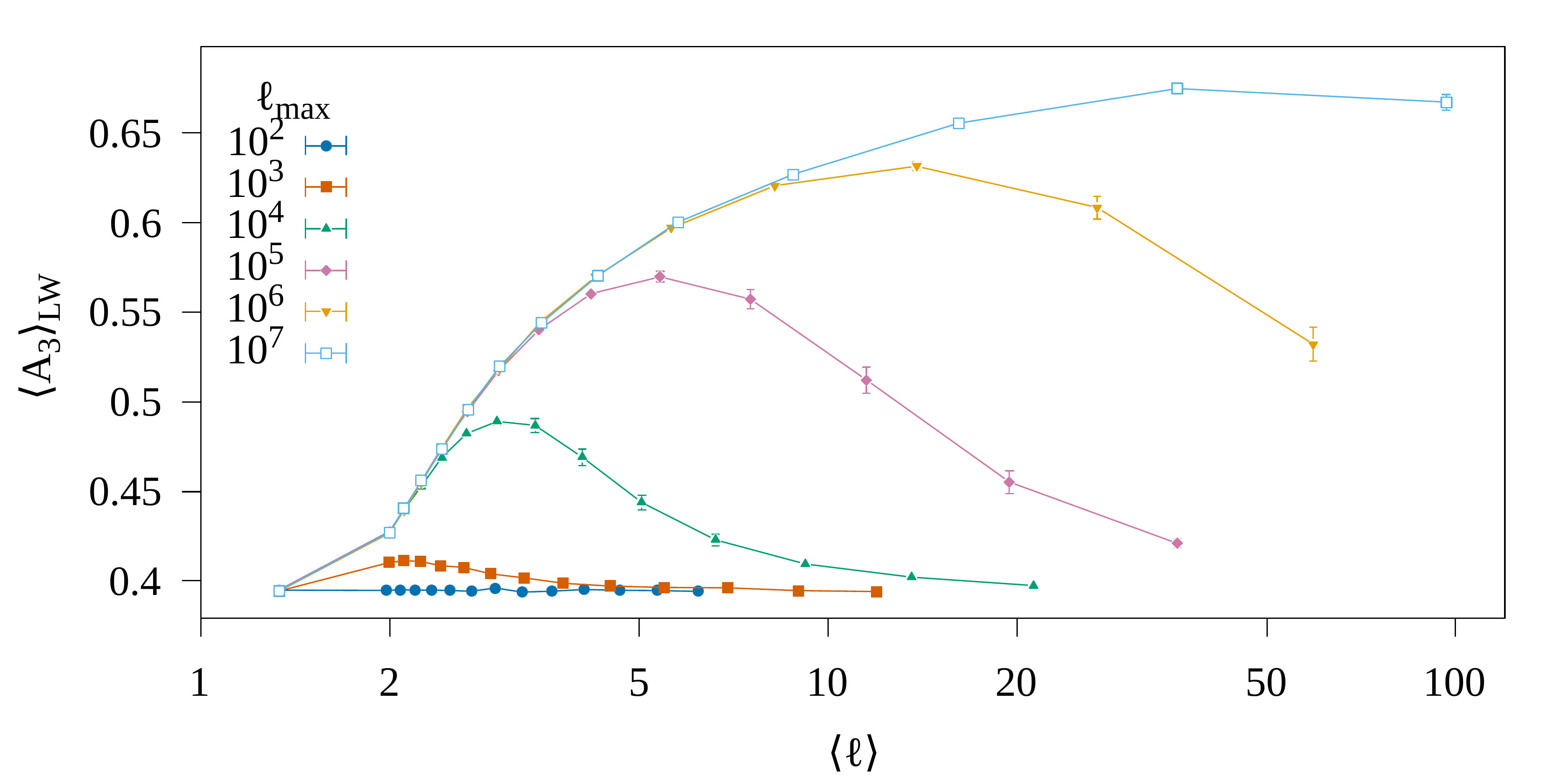}
    \includegraphics[width=\linewidth]{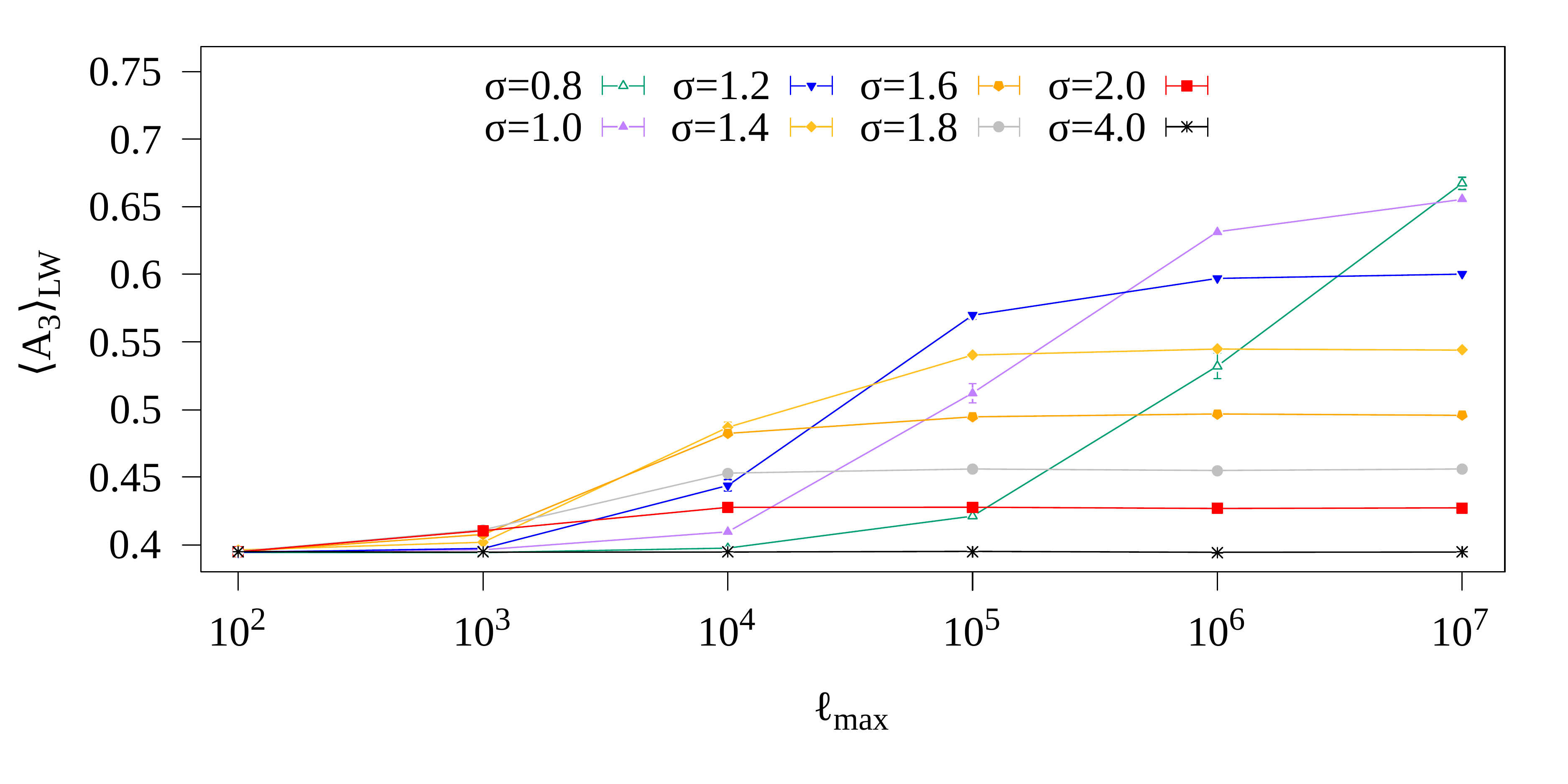}
    \caption{Average asphericity $\langle A_3\rangle_{{\rm LW}}$ as a
function of $\sigma$ (upper panel), $\langle\ell\rangle$ (middle panel)  at fixed $\ell_{\max}$, and as a function of $\ell_{\max}$ at fixed $\sigma$ (lower panel). For each parameter pair, the $N$-dependent ensemble means are further averaged over $70000\le N\le10^5$. Error bars show their standard deviation across this interval. Lines connect the points as eye guides.}
    \label{fig:A-compare}
\end{figure}

\begin{figure}
    \centering
    \includegraphics[width=\linewidth]{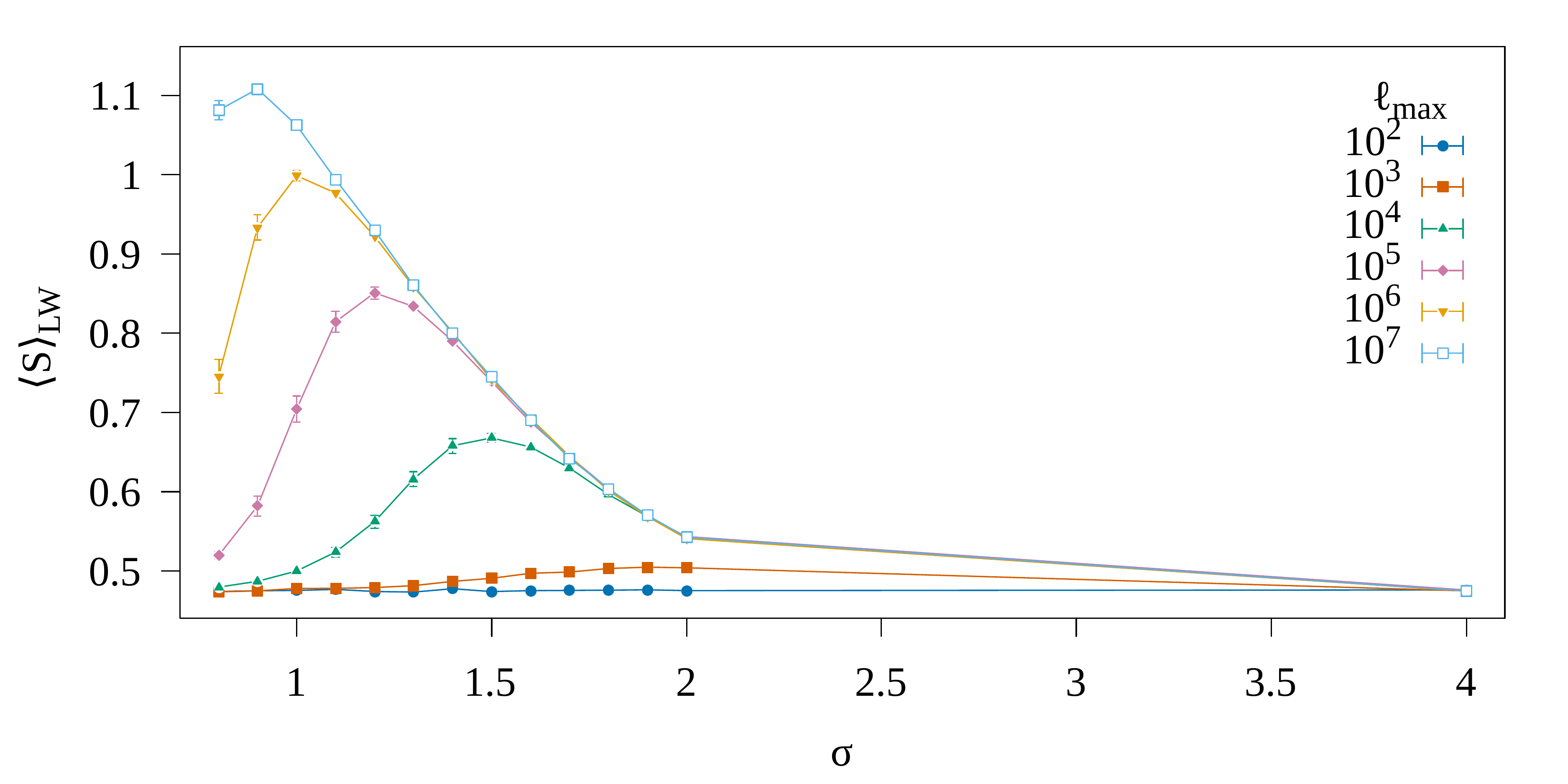}
    \includegraphics[width=\linewidth]{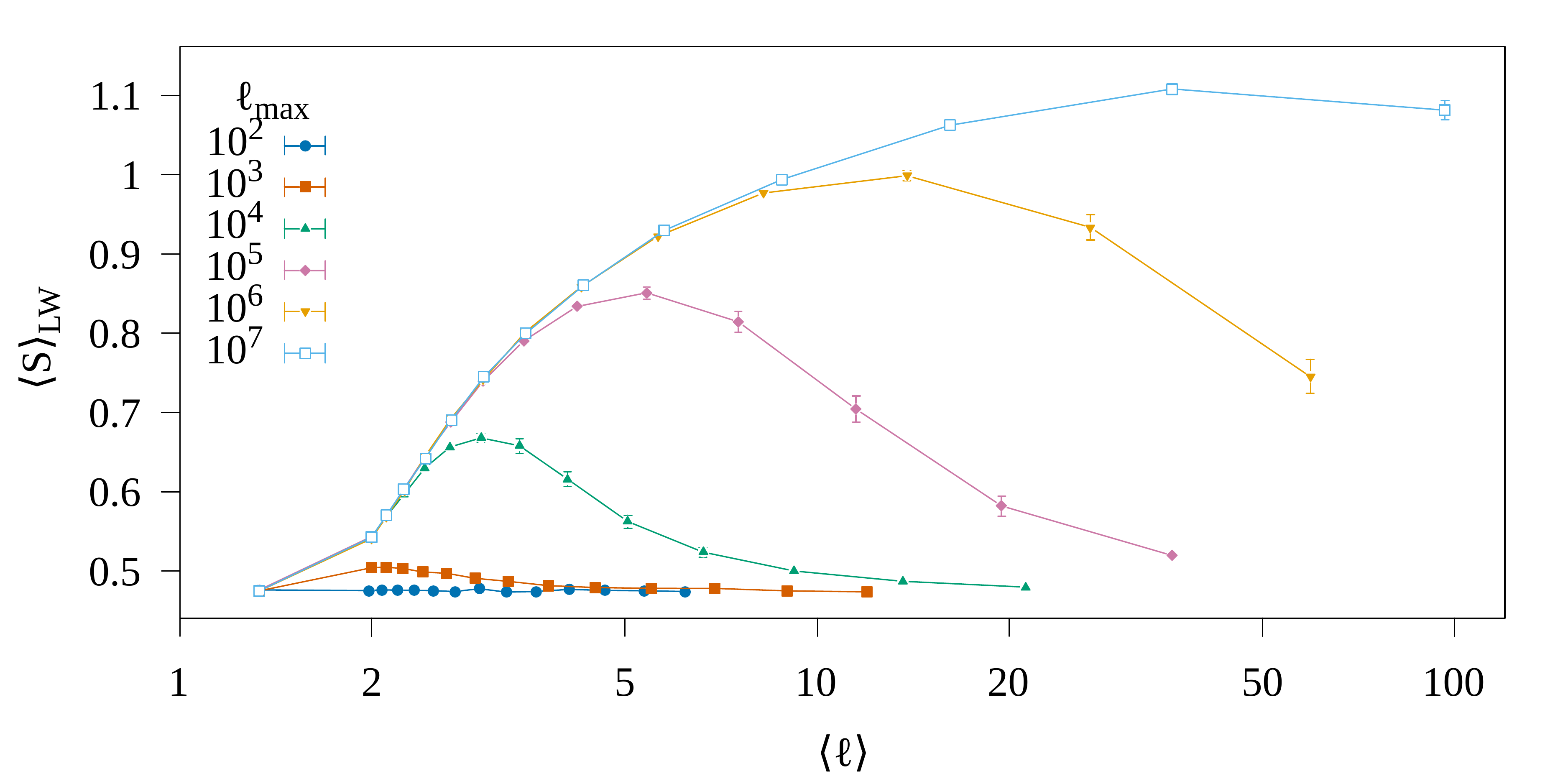}
    \includegraphics[width=\linewidth]{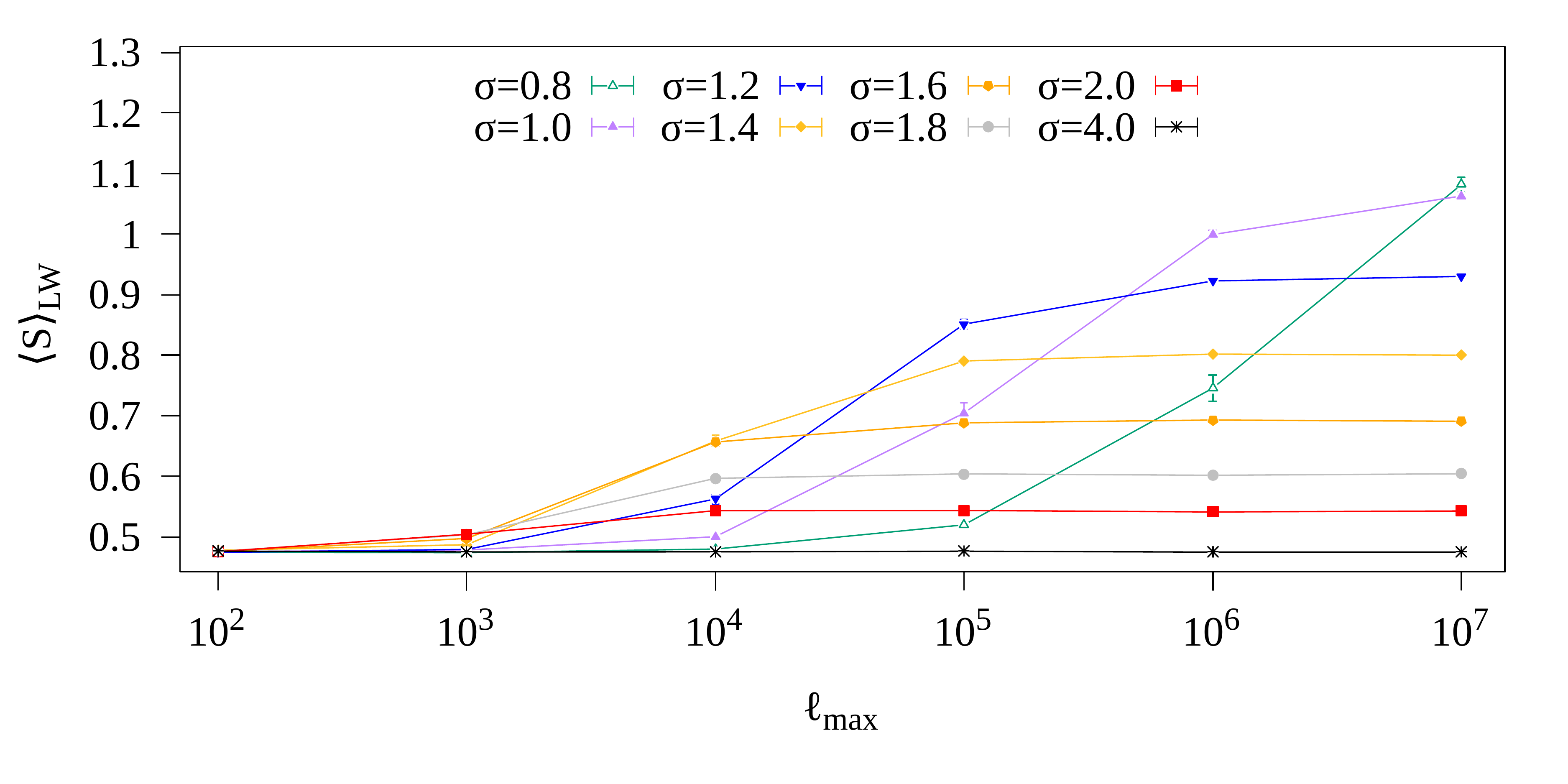}
    \caption{Average prolateness $\langle S\rangle_{{\rm LW}}$ as a
function of $\sigma$ (upper panel), $\langle\ell\rangle$ (middle panel) at fixed $\ell_{\max}$, and as a function of $\ell_{\max}$ at fixed $\sigma$ (lower panel). For each parameter pair, the $N$-dependent ensemble means are further averaged over $70000\le N\le10^5$. Error bars show their standard deviation across this interval. Lines connect the points as eye guides.}
    \label{fig:S-compare}
\end{figure}

The upper panels of Figs.~\ref{fig:A-compare} and~\ref{fig:S-compare} show a non-monotonic variation of both shape parameters with $\sigma$. For the smallest cutoff, the curves are nearly flat and stay close to the Gaussian-chain reference values. As $\ell_{\max}$ grows, clear peaks emerge in both mean asphericity and mean prolateness, shifting toward smaller $\sigma$ with increasing cutoff. Across the sampled parameters, these maxima are located near $\sigma \approx 1.5$ for $\ell_{\max}=10^4$ and near $\sigma \approx 1.0$ for $\ell_{\max}=10^6$. Hence, a heavier step-length distribution tail does not automatically yield stronger average shape anisotropy within this range.

The lower panels of Figs.~\ref{fig:A-compare} and~\ref{fig:S-compare} highlight the cutoff dependence at fixed $\sigma$. At $\sigma=1.4$, both observables show little variation once $\ell_{\max}$ reaches $10^5$--$10^6$. By contrast, the data for $\sigma=0.8$ and $1.0$ remain sensitive to the cutoff up through $\ell_{\max}=10^7$. At $\sigma=4.0$, both shape parameters stay near their Gaussian limits across all cutoff values. Although reflecting finite chain lengths, the leveling off observed in several curves hints at the $N\to\infty$ asymptotic behavior.

The middle panels of Figs.~\ref{fig:A-compare} and~\ref{fig:S-compare} present the same shape parameters as functions of the measured mean segment length $\langle\ell\rangle$. This representation retains a strong dependence on $\ell_{\max}$: comparable mean segment lengths can correspond to substantially different asphericities and prolatenesses. Within the investigated chain-length and cutoff ranges, $\langle\ell\rangle$ alone is insufficient to determine the mean shape characteristics. The full distribution must also be taken into account.

\subsection{L\'evy walks. Probability distributions}
\begin{figure}
\includegraphics[width=\linewidth]{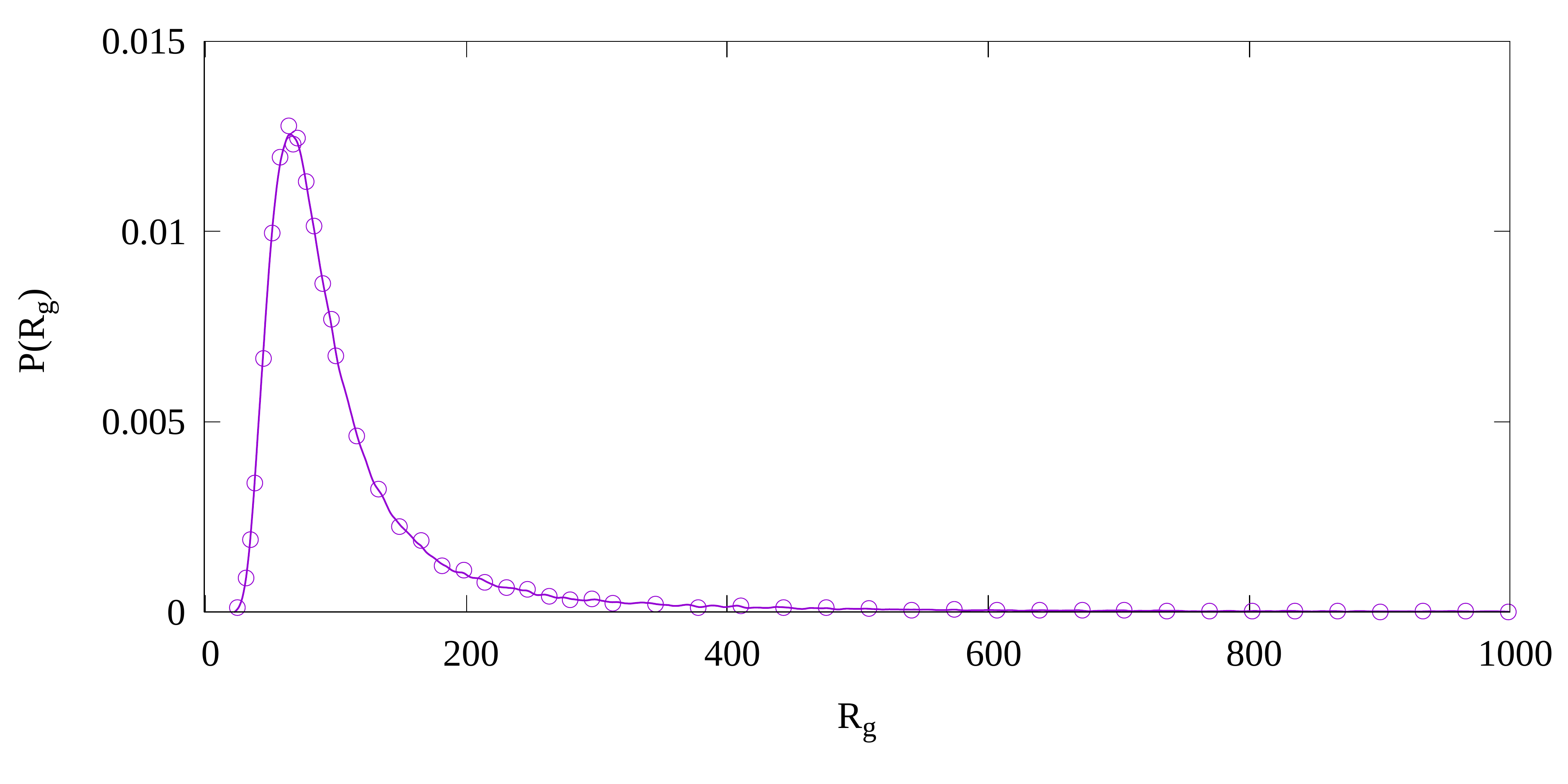}
\caption{Probability density of the radius of gyration $R_{{\rm g}}$ of LW trajectories for
$\sigma=1.5$, $N=10^5$, $M=10^5$, and $\ell_{\max}=10^5$.} \label{fig:histogram-rg}
\end{figure}

The distributions of the observables provide
information that is not contained in their ensemble averages alone.
As shown in Fig.~\ref{fig:histogram-rg}, the distribution of $R_{{\rm g}}$ is asymmetric and
possesses a pronounced tail towards large values, reflecting the occurrence of configurations containing unusually long segments.

\begin{figure}
\includegraphics[width=\linewidth]{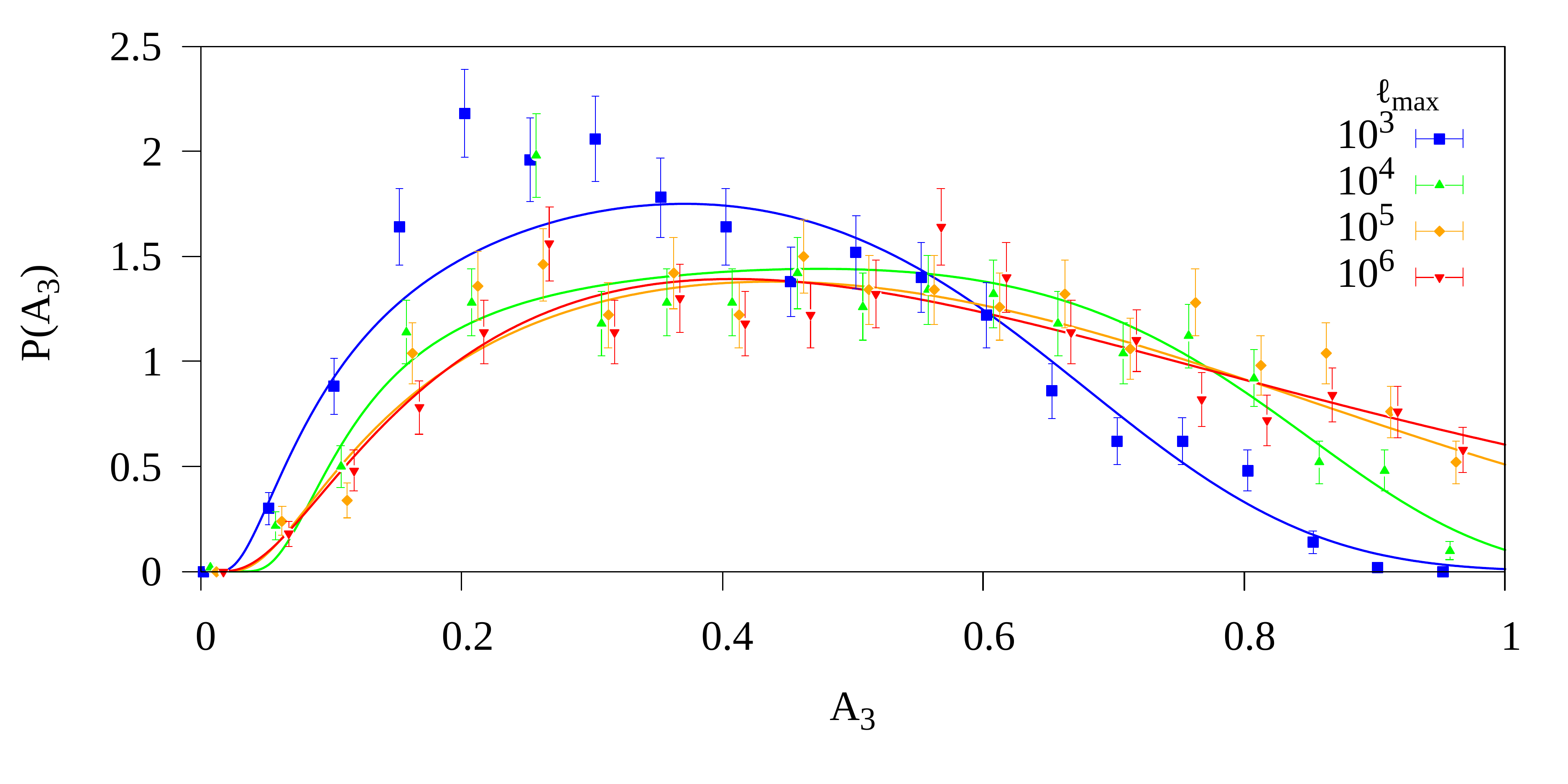}
\includegraphics[width=\linewidth]{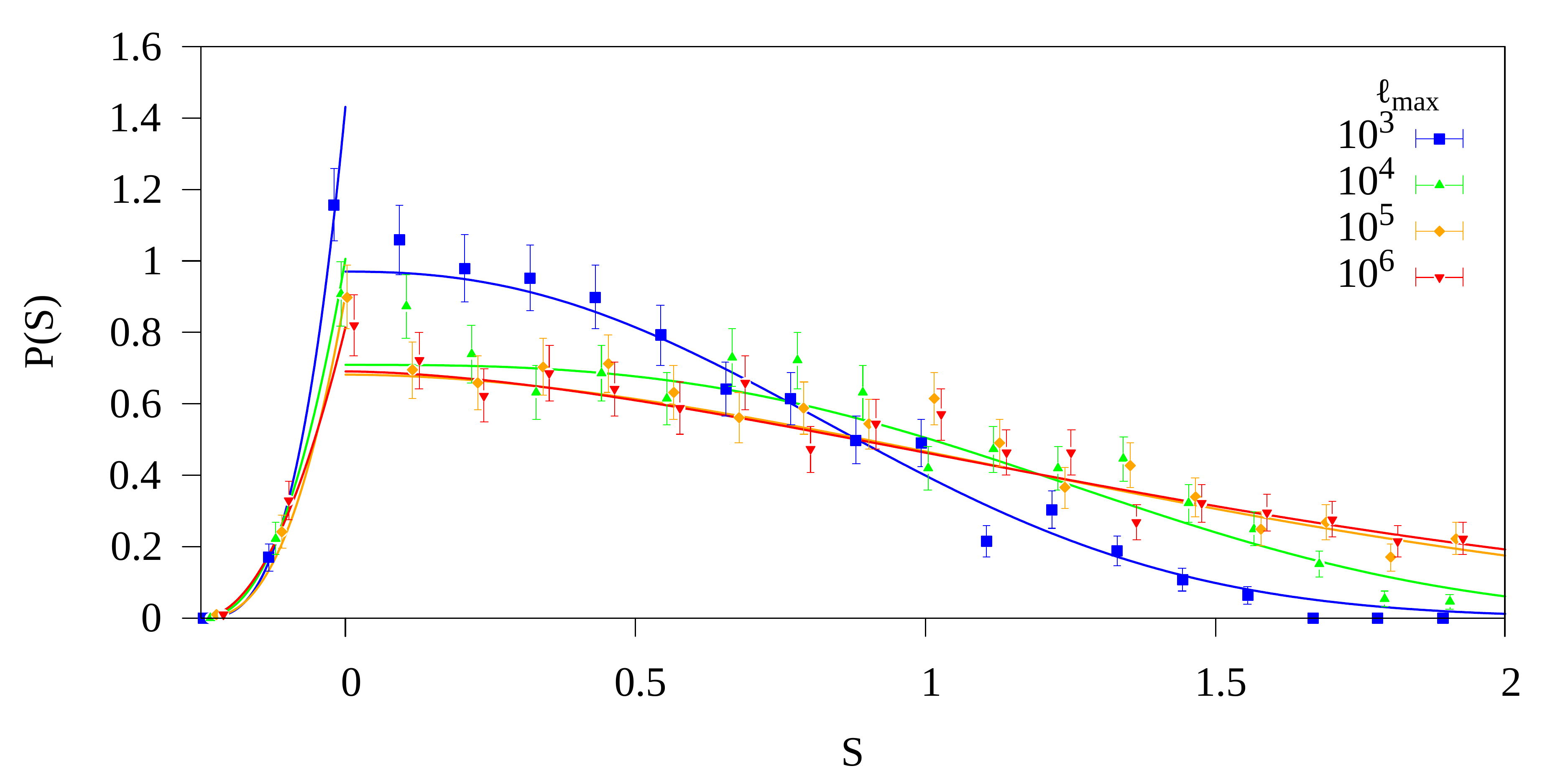}
\caption{Probability densities of the asphericity $A_3$ (above) and
prolateness $S$ (below) of LW trajectories for $\sigma=1.5$, $N=10^5$, $M=10^5$, and different $\ell_{\max}$ from $10^3$ to $10^6$. Symbols with error bars represent histogram estimates.
Solid lines show empirical fits using an extended Lhuillier form for $A_3$, and separate power-law ($S<0$) and stretched-exponential ($S>0$) forms for $S$.}\label{fig:histogram-shape}
\end{figure}

Much broader distributions are observed for the shape parameters, as presented in Fig.~\ref{fig:histogram-shape}. The asphericity distribution has most of its statistical weight away from $A_3=0$, showing that nearly spherical configurations are uncommon. The prolateness distribution is predominantly concentrated at positive values of $S$, indicating that prolate configurations occur more frequently than oblate ones.
The broadness of both distributions demonstrates that the mean values $\langle A_3\rangle$ and $\langle S\rangle$ represent ensembles containing configurations with notably different geometries.

To describe the distributions in Fig.~\ref{fig:histogram-shape} quantitatively, we use the empirical fitting forms employed in Ref.~\cite{Kalyuzhnyi2016}. For the asphericity, we adopt the extended Lhuillier-like expression
\begin{equation}
    P(A_3)\simeq C_A\exp\left[
    -\left(\frac{a}{A_3}\right)^{\alpha}
    -\left(\frac{A_3}{b}\right)^{\beta}
    \right],
    \label{eq:asphericity-distribution-fit}
\end{equation}
where $C_A$, $a$, $b$, $\alpha$, and $\beta$ are fitting parameters. This form extends an ansatz originally developed for polymer size distributions \cite{Lhuillier1988}.

For the prolateness, the branches $S<0$ and $S>0$ are fitted separately by $C_-(S+1/4)^{\eta}$ and
$C_+\exp[-(S/S_0)^{\gamma}]$, respectively \cite{Kalyuzhnyi2016}. These empirical fits provide compact descriptions of the distribution shapes and allow comparison between different cutoffs beyond their means. 

The fitted distributions can also be used to estimate mean values from their normalized first moments:
\begin{equation}
    \langle X\rangle_{{\rm fit}} =
    \frac{\int_{\Omega_X} x\,f_X(x)\,dx}
         {\int_{\Omega_X} f_X(x)\,dx},
    \qquad X\in\{A_3,S\},
\end{equation}
where $f_X$ denotes the fitted function,
$\Omega_{A_3}=[0,1]$, and $\Omega_S=[-1/4,2]$.
Such estimates may provide a complementary consistency check.

\section{Conclusions} \label{sec:conclusions}

The concept of scaling is central to polymer physics, with origins in the foundational works of Kuhn, Flory, Huggins and others. Building on the work of de Gennes and des Cloizeaux \cite{deGennes79,desCloiseaux1991}, the scaling behavior of long, flexible macromolecules in a good solvent was linked to critical phenomena near continuous phase transitions, establishing the basis for applying field-theoretic methods to polymer physics. The random walk and self-avoiding walk models serve as the standard paradigm for modeling flexible polymer chains. The objective of our work is to introduce and test a similar paradigm for describing semiflexible polymers; to this end, we examine two models:
the persistent random walks (PRW) and truncated L\'evy walks (LW). Within the PRW model,  the probability $p$ is introduced to enhance the trajectory local stiffness, see Eq. (\ref{Ppers}). Accordingly, the persistence length $l_{\rm p}$  gradually increases with $p$,  Eq. (\ref{kpers}). Within the LW model, the mean step length of a L\'evy distribution, Eq. \eqref{PLevy}, is treated as an effective persistence length $l_{\rm p}$  of a semiflexible polymer chain. 

To test these models, in addition to calculating scaling parameters—specifically, the scaling of the mean-square distance—we also consider universal shape properties, such as asphericity and prolateness, which are essential for analyzing the rheological properties of polymer solutions.
It is worth noting that, just as polymer scaling exponents are related to the universal critical exponents of continuous phase transitions, 
polymer shape parameters are related to universal combinations of critical amplitudes.
We focused on the impact of the statistical properties of the segment-length distributions (\ref{Ppers}) and (\ref{PLevy})  on the universal size and shape characteristics of semiflexible polymer chains. For the PRW model, we confirmed a crossover governed by the persistence length scale: the trajectory exhibits  the Gaussian regime with scaling exponent $\nu = 1/2$ till the immediate vicinity of $p \to 1$, where  transition to the rigid-rod regime with $\nu = 1$ occurs abruptly. The same trend was observed for the
 shape parameters, which are attaining the Gaussian values of a flexible polymer chain ($\langle A_{3}^ {\rm RW}\rangle \approx0.394$ \cite{Wei1997a}, and
$\langle S^{{\rm RW}}\rangle \approx0.475$ \cite{Jagodzinski1992}) until the abrupt chain stretching at $p\to1$  into strongly elongated rigid-rod conformation with $\langle A_{3} \rangle=1$ and
$\langle S\rangle =2$.
On the other hand, in the case of LW model the observed behavior is much more rich. Decreasing the L\'evy exponent $\sigma $  leads to an increase in the numerical estimates of effective scaling exponent $\nu $  from a near-Gaussian value at $\sigma>2$ up to superdiffusive values, merging towards the theoretical reference $\nu=1/\sigma$ at sufficiently large values of upper cutoff $\ell_{{\rm max}}$. The crucial role of the parameter $\ell_{{\rm max}}$ is observed: the extremely long L\'evy jumps, determining the superdiffusion
nature of the motion  are artificially truncated as  $\sigma$ increases,
which leads to an ``underestimation'' of the exponent $\nu$ for $\sigma<1.4$. 
The shape parameters exhibit a gradual increase with decreasing $\sigma$ (increasing the mean length of straight segment $l_p$ of LW trajectory).
From this perspective, the LW model confirms the continuous change in the shape parameters $\langle A \rangle$ and $\langle S \rangle$  of semiflexible macromolecules with an increasing persistence length, as observed in Refs. \cite{Alim2007,Chen2020,Fonari2013,Liu2017}.

 The results obtained support the potential of application of the L\'evy-walk models for capturing the geometric size and shape  properties of semi-flexible polymers.
In particular, we demonstrate that the extended structural segments can fundamentally impact the global shape anisotropy and folding kinetics of semiflexible polymer chains. On the other hand, the constraints connected with a finite cutoff $\ell _{{\rm max}}$ and finite chain length should be carefully taken into account when interpreting simulations within the LW framework.
While the obtained values of the universal scaling exponent are compatible with theoretical predictions and indicate a 
shift in the universality class, our results for the shape parameters $\langle A_3 \rangle$ and $\langle S \rangle$ are, to the best of our knowledge, the first 
numerical observations of their kind and call for further analytical confirmation.

\section*{Acknowledgments}
We thank Khrystyna Haydukivska, Jaroslav Ilnytskyi and Dmytro
Yaremchuk for useful discussions.
VB and YulH acknowledge support from the National Research Foundation of Ukraine under Project No. 2025.07/0366, ``Functional efficiency of complex multibranched polymers in adsorption-based applications: computer simulations and predictions." 

\bibliography{Levypapers}

\end{document}